\documentclass[fleqn,usenatbib,usedcolumn]{mnras}

\usepackage[T1]{fontenc}
\usepackage{ae,aecompl}

\usepackage{multirow}
\usepackage{caption}
\usepackage{dsfont}
\usepackage{xcolor}
\usepackage{arydshln}
\usepackage{commath}

\usepackage{graphicx}
\usepackage{bbold}
\usepackage{graphicx}	% Including figure files
\usepackage{amsmath}	% Advanced maths commands
\usepackage{amssymb}	% Extra maths symbols
\usepackage{csquotes}
\usepackage{anyfontsize}
\usepackage{physics}
\usepackage{newtxtext,newtxmath}
\usepackage[capitalise]{cleveref}
\newcommand{\noop}[1]{}

\newcommand\pcref[1]{(Eq.~\ref{#1})}
\title[SP(k) - A baryon suppression model of $P(k)$]{SP(k) - A hydrodynamical simulation-based model for the impact of baryon physics on the non-linear matter power spectrum}

\author[Salcido et al.]{Jaime Salcido$^{\href{https://orcid.org/0000-0002-8918-5229}{\includegraphics[scale=0.04]{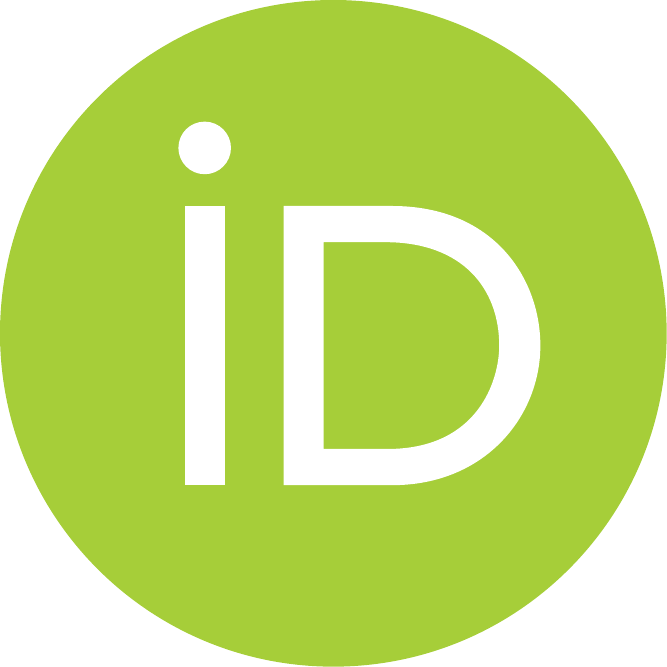}} 1}$\thanks{E-mail: \href{mailto:j.salcidonegrete@ljmu.ac.uk}{j.salcidonegrete@ljmu.ac.uk}}, 
Ian G. McCarthy$^{\href{https://orcid.org/0000-0002-1286-483X}{\includegraphics[scale=0.04]{ORCIDiD.pdf}} 1}$\thanks{E-mail:
\href{mailto:i.g.mccarthy@ljmu.ac.uk}{i.g.mccarthy@ljmu.ac.uk}}, 
Juliana Kwan$^{1}$,
Amol Upadhye$^{\href{https://orcid.org/0000-0003-1345-8224}{\includegraphics[scale=0.04]{ORCIDiD.pdf}} 1}$,
Andreea S. Font$^{\href{https://orcid.org/0000-0001-8405-9883}{\includegraphics[scale=0.04]{ORCIDiD.pdf}} 1}$
\\ \\
$^{1}$ Astrophysics Research Institute, Liverpool John Moores University, 146 Brownlow Hill, Liverpool L3 5RF, UK}

\date{Accepted XXX. Received YYY; in original form ZZZ}

\pubyear{2020}

\volume{{\rm in press}}

\begin{document}
\label{firstpage}
\pagerange{\pageref{firstpage}--\pageref{lastpage}}
\maketitle

\begin{abstract}
Upcoming large-scale structure surveys will measure the matter power spectrum to approximately percent level accuracy with the aim of searching for evidence for new physics beyond the standard model of cosmology. In order to avoid biasing our conclusions, the theoretical predictions need to be at least as accurate as the measurements for a given choice of cosmological parameters. However, recent theoretical work has shown that complex physical processes associated with galaxy formation (particularly energetic feedback processes associated with stars and especially supermassive black holes) can alter the predictions by many times larger than the required accuracy. Here we present \texttt{SP(k)}, a model for the effects of baryon physics on the non-linear matter power spectrum based on a new large suite of hydrodynamical simulations. Specifically, the ANTILLES suite consists of 400 simulations spanning a very wide range of the ``feedback landscape'' and show that the effects of baryons on the matter power spectrum can be understood at approaching the percent level in terms of the mean baryon fraction of haloes, at scales of up to {$k \, {\lesssim} \, 10 h$ Mpc$^{-1}$} and redshifts up to $z=3$. For the range of scales and redshifts that will be probed by forthcoming cosmic shear measurements, most of the effects are driven by galaxy group-mass haloes ($M \sim 10^{13-14}$ M$_\odot$). We present a simple Python implementation of our model, available at \href{https://github.com/jemme07/pyspk}{https://github.com/jemme07/pyspk}, which can be used to incorporate baryon effects in standard gravity-only predictions, allowing for marginalisation over baryon physics within cosmological pipelines.
\end{abstract}


%%%%%%%%%%%%%%%%%%%%%%%%%%%%%%%%%%%%%%%%%%%%%%%%%%

%%%%%%%%%%%%%%%%% BODY OF PAPER %%%%%%%%%%%%%%%%%%

\section{Introduction}\label{sec:intro}

Measurements of the growth of large-scale structure (LSS) provide an important test of our cosmological theoretical framework \citep{peebles_large-scale_1980,Bond_80,davis_evolution_1985, kaiser_clustering_1987,Peacock_94}.  They are independent of, and complementary to, constraints from analyses of fluctuations in the cosmic microwave background (CMB) and geometric probes, such as Type Ia supernovae (SNe) and baryon acoustic oscillations (BAOs).  The different LSS tests (e.g., galaxy clustering, Sunyaev-Zel'dovich power spectrum, cosmic shear, CMB lensing, redshift-space distortions, etc.) characterise the clustering of matter on different scales and its evolution over cosmic time.  On the largest scales where the fluctuations are small and gravity is the sole force of relevance, perturbation theory is sufficiently accurate to calculate the clustering of matter.  However, at present most LSS tests probe well into the non-linear regime, since that is typically where most of the observational signal and much of the cosmological sensitivity originates (e.g., \citealt{amon_2022}).  The standard approach is therefore to adopt the (semi-)analytic `halo model' formalism \citep{peacock_halo_2000,seljak_analytic_2000,ma_2000,cooray_halo_2002,smith_stable_2003,Mead_16,acuto_2021}, which is often calibrated using aspects of large N-body cosmological simulations, or to use such N-body simulations to directly correct linear theory in an empirical fashion (e.g., \citealt{Takahashi_12,heitmann_2016,lawrence_2017,derose_2019,euclid_2019,angulo_2021}).

These approaches would be fully adequate if the matter in the universe were composed entirely of dark matter. However, baryons contribute a non-negligible fraction of the matter density ($\Omega_b/\Omega_m \approx 0.157 \pm 0.001$; \citealt{planck_2018_cosmology}) and work based on cosmological \textit{hydrodynamical} simulations has shown that feedback processes associated with galaxy formation can have a relatively large effect (typically many times larger than the statistical precision of upcoming surveys) on the matter distribution on scales of up to a few tens of megaparsecs \citep{van_daalen_effects_2011,schneider_new_2015,mummery_separate_2017,springel_first_2018,van_daalen_2020}. It is therefore crucially important that we model these effects as accurately as possible, as they will introduce significant biases in the inferred cosmological parameters from upcoming surveys if no action is taken (e.g., \citealt{semboloni_quantifying_2011,schneider_2020,castro_2021,debackere_2021}).

One way to tackle this challenging problem is to modify the gravity-only predictions with simple analytic prescriptions for baryon physics that have some number of associated free parameters and to jointly constrain the cosmological and feedback parameters through comparisons to LSS observables. Examples of this approach include the halo model \textsc{HMcode} of \citet{Mead_16} (see also \citealt{mead_hmcode-2020_2020}) or the `baryonification' method of \citet{schneider_new_2015} and \citet{arico_2021}, which directly modifies the outputs of gravity-only simulations by adjusting the radial position of particles within haloes in order to mimic the effects of baryons on small scales. Some of the strengths of these approaches include: i) they are considerably computationally cheaper than running full cosmological hydrodynamical simulations; ii) the baryon prescriptions are generally flexible and easy to adjust; and iii) they can be straightforwardly incorporated within existing pipelines based on gravity-only simulations or the halo model.

Recent cosmic shear surveys have employed these methods in an attempt to account for and measure, baryonic effects.  For example, the fiducial Kilo Degree Survey (KiDS) analysis employs a halo model prescription where the effects of baryons are incorporated by marginalising over a phenomenological `bloating' parameter that modifies the concentrations of dark matter haloes (e.g., \citealt{Asgari2021, Heymans2021, Troster2021}). The Dark Energy Survey (DES) have taken a different approach, by introducing scale cuts to remove small-scale measurements that are most affected by baryonic effects (e.g., \citealt{derose_2019,Amon2022,Krause2021arXiv,Secco2022}). This approach is motivated by the concern that errors in the modeling of baryonic effects could lead to biased cosmological parameter estimates. However, introducing scale cuts comes at the expense of losing valuable information on small scales, which can be important for constraining cosmological parameters and testing extensions to the standard model. More recently, the full DES data (including small-scale measurements) has been reanalysed using the baryonification approach to account for the impact of baryon physics \citep{chen_2023,arico_2023}.  Both the re-analysis of the DES data and a recent joint analysis of KiDS cosmic shear and Sunyaev-Zel'dovich effect data \citep{troster_2022} using a more physical halo model \citep{mead_hydrodynamical_2020}, detect the impact of baryon physics even in current data, at the approximately $2-3 \, \sigma$ level.  These studies conclude that the implied level of suppression of the power spectrum is consistent with the predictions of recent calibrated hydrodynamical simulations such as BAHAMAS (see below).

While the halo model and baryonification approaches have some important strengths, there are also some disadvantages to these methods.  Particularly, the modelling of baryon physics and its back reaction on dark matter is simplistic and generally not self-consistent and that there may be non-negligible degeneracies between the various baryonic `nuisance' parameters themselves and also between those parameters and the cosmological parameters being varied. Marginalisation over the uncertain baryon physics may therefore lead to a significant degradation of the cosmological constraining power.  Of course, cosmological hydrodynamical simulations also have adjustable free parameters (as discussed immediately below), but the resulting diversity of outcomes in terms of the matter clustering is likely to be more constrained in hydrodynamical simulations than simple empirical models would allow.  For example, the baryon fraction--halo mass relation cannot be arbitrarily steep in simulations because high-mass objects are assembled from the accretion of lower-mass objects (e.g., \citealt{balogh_2008}) and haloes can recapture gas as they grow, too.  Also, naturally emerging conditions such as convective and virial equilibrium place constraints on the radial distribution of matter within haloes, and so on.  The upshot is that we expect many of the parameters that characterise mass distributions of groups and clusters to be physically correlated rather than independent of each other.

Full cosmological hydrodynamical simulations could therefore be employed as a means of self-consistently incorporating the impact of baryons on LSS when constraining cosmology.  However, this has so far not proved possible due to their computational expense, noting that a typical MCMC chain in cosmology can require $\sim10^5$ evaluations whereas currently available suites of simulations typically only contain a handful of realisations.  Furthermore, a major obstacle in directly simulating the impact of baryon physics on the matter clustering is that current simulations do not resolve all of the physical scales necessary to capture such processes in an ab initio way.  Consequently, so-called subgrid models are required to include these effects and they often have considerable uncertainties, which are not unlike the uncertainties in simple baryon models discussed above (although the impact of subgrid models is often bounded by physical constraints, whereas the empirical models may not be, as discussed above).  Indeed, previous simulation work has shown that variations of the parameters associated with the efficiencies of feedback processes even within plausible bounds can lead to relatively large differences in the predicted properties of galaxy groups and clusters (e.g., \citealt{puchwein_2008,mccarthy_case_2010,le_brun_towards_2014,planelles_2014,oppenheimer_2021}) which dominate the matter clustering \citep{van_daalen_contributions_2015,mead_hydrodynamical_2020}.  A consequence of these variations in the predicted properties of groups/clusters is relatively large study-to-study variations in the predicted impact of baryons on the matter power spectrum (see the simulation comparisons in \citealt{chisari_modelling_2019} and \citealt{van_daalen_2020}), in spite of the simulations being more constrained than phenomenological models.

As discussed recently in \citet{oppenheimer_2021}, the variations in the predicted properties of groups/clusters and the impact of baryons on the matter power spectrum from hydrodynamical simulations in the literature is not unexpected.  It is a consequence of not being able to derive the efficiencies for the relevant feedback processes from first principles (see discussion in \citealt{schaye_eagle_2015}).  The efficiency of feedback in simulations must therefore generally be calibrated in order to ensure they reproduce particular observed quantities, after which the realism of the simulations may be tested against independent quantities.  The approach of the \textsc{BAHAMAS} programme \citep{mccarthy_bahamas_2017,mccarthy_bahamas_2018} (see also the recent FABLE simulations; \citealt{henden_2018,henden_2020}), was to explicitly calibrate the feedback efficiencies so that they reproduce the observed baryon fractions of galaxy groups.  Aside from an explicit dependence on the universal baryon fraction \citep{white_1993}, $\Omega_b/\Omega_m$, which is tightly constrained by the CMB, the baryon fractions of groups should be insensitive to changes in cosmology and therefore represents a fairly ideal quantity on which to calibrate the feedback.  Note also that since the growth of fluctuations is fundamentally a gravitational process, by ensuring the simulations have the correct baryon fractions on the scale of groups/clusters, the impact of baryons on $P(k)$ ought to be strongly constrained by this approach.  \citet{van_daalen_2020} have recently confirmed this simple picture by demonstrating that the differences in the predicted impact of baryons on the present-day $P(k)$ from different simulations can be understood at approximately the percent level up to $k\approx1$ $h$ Mpc$^{-1}$ in terms of the differences in baryon fraction in the various simulations at a mass scale of $\sim10^{14}$ M$_\odot$.   We highlight here that the simulations analysed in that study varied by more than a factor of a thousand in mass resolution, used different hydro solvers and subgrid physics implementations, assumed different baseline cosmologies, and some (specifically \textsc{BAHAMAS}) varied the initial conditions to explore the potential impact of cosmic variance.  None of these variations were found to significantly affect the impact of baryons on $P(k)$ after differences due to the group baryon fractions were factored out.

The results of \citet{van_daalen_2020} are very promising and potentially offer a path forward for incorporating the impact of baryons on the matter power spectrum from cosmological hydrodynamical simulations. Note that with a mapping between the baryon fraction and the impact of baryons $P(k)$, there is no longer a necessity to calibrate the hydrodynamical simulations to high precision to match some particular data set. Instead, one can properly account for the uncertainties in the feedback/subgrid effects on $P(k)$ by, for example, conservatively marginalising over the uncertainties in the observed baryon fractions. Furthermore, with a simple mapping, the correction to the gravity-only clustering could be straightforwardly included in cosmological pipelines because it can be rapidly evaluated.  However, before these aims can be achieved a number of limitations must first be overcome. First and foremost, the quantitative link between halo baryon fraction and the suppression of the power spectrum must be established for a wider range of models. \citet{van_daalen_2020} used a small number ($\approx10$) of publicly-available simulations which likely do not bracket the full range of possible behaviours. Furthermore, the mapping between baryon fraction and $P(k)$ was established only at $z=0$ and up to a maximum wavenumber of $k\approx1$ $h$ Mpc$^{-1}$, both of which are insufficient for current and future LSS tests.

In the present study, we overcome these limitations by presenting a new large suite of cosmological hydrodynamical simulations designed specifically to build a mathematical model for the suppression of the matter power spectrum with the baryon fractions of galaxy groups as its input. The ANTILLES suite contains 400 cosmological hydrodynamical simulations that vary both the important parameters that characterise the efficiencies of stellar and active galactic nuclei (AGN) feedback in the simulations as well as the employed hydrodynamics scheme, and span an unprecedentedly wide range of behaviours in terms of baryon fractions and $P(k)$ modifications. We develop a model that can reproduce the effects of baryons on $P(k)$ to typically better than ${\approx}2\%$ precision up to $k = 10$ $h$ Mpc$^{-1}$ out to a redshift of $z=3$.

The present study is structured as follows. In Section \ref{sec:Sim} we describe the new suite of cosmological hydrodynamical simulations. In Section \ref{sec:model} we present \texttt{SP(k)}, an empirical model that provides the mapping between the observable baryon fractions of groups/clusters and the suppression of the matter power spectrum, $P(k)$. In Section \ref{sec:bahamas} we test \texttt{SP(k)} against the independent \textsc{BAHAMAS} simulations. In Section \ref{sec:discussion} we discuss some limitations of the present work and in Section \ref{sec:conclusion} we summarise our findings.

\section{Simulations}\label{sec:Sim}

To quantify the potential impact of baryon physics on the non-linear matter power spectrum, our simulations must satisfy a number of requirements. Firstly, the simulated volumes must be sufficiently large to contain a representative sample of the haloes that contribute most significantly to the matter power spectrum. They must also be of sufficiently high resolution to resolve the range of scales over which cosmological measurements are made and so that we resolve the locations of important feedback processes (specifically, the simulations contain the haloes from which the majority of the energetic feedback originates). Finally, we need to explore a wide range of feedback possibilities, which we dub the ``feedback landscape'', such that key properties such as the baryon fractions (stellar and gas fractions) span a wide range and conservatively bracket current and hopefully future observational constraints.

With these requirements in mind, we have created a new bespoke suite of 400 simulations, the ANTILLES suite. Each simulation has a box size of 100 Mpc/$h$ on a side, which \citet{van_daalen_2020} have shown is sufficiently large to capture the relative effects of baryons on the matter power spectrum (see their appendix A).  Our simulations adopt the same mass resolution as BAHAMAS, which \citet{mccarthy_bahamas_2017,mccarthy_bahamas_2018} have shown is sufficiently high for various large-scale structure tests (see also appendix A of \citealt{van_daalen_2020}). Thus, each simulation has $256^3$ baryon and dark matter particles (each), corresponding to (initial) masses\footnote{Unless otherwise stated, throughout the paper masses are specified in $\mathrm{M}_\odot$ (not $h^{-1} \mathrm{M}_\odot$).} of $m_\mathrm{g} = 1.09 \times 10^9 \mathrm{M}_\odot$ and $m_\mathrm{dm}=5.51 \times 10^9 \mathrm{M}_\odot$ respectively, given our choice of cosmology (below). The gravitational softening is fixed to $4 h^{-1}$ kpc in physical coordinates below $z = 3$ and in comoving coordinates at higher redshifts.  

We adopt a flat $\Lambda$CDM cosmology consistent with the WMAP 9-year results \citep{2013ApJS..208...19H}. The cosmological parameters used in these simulations are listed in \cref{tab:cosmo_params}.  As we focus on the \textit{relative} impact of baryons on the matter power spectrum, we do not expect the precise choice of cosmology to be important for our purposes.  We nevertheless discuss the possible dependence of our results on cosmology in Section \ref{sec:discussion}.

\begin{table}
\centering
\caption{The cosmological parameters for the simulations used in this study. We adopt a flat $\Lambda$CDM cosmology with WMAP 9-year based cosmological parameters \citep{2013ApJS..208...19H}. $\Omega_\mathrm{m}, \Omega_\Lambda, \Omega_\mathrm{b}$ are the average densities of matter, dark energy, and baryonic matter in units of the critical density at redshift $z=0$; $H_0$ is the Hubble constant, $\sigma_8$ is the square root of the linear variance of the matter distribution when smoothed with a top-hat filter of radius $8 \, h^{-1}$ cMpc, and $n_s$ is the scalar power-law index of the power spectrum of primordial adiabatic perturbations.} \label{tab:cosmo_params}
\begin{tabular}{ll}
\hline
Cosmological Parameter                                                    & Value               \\ \hline
$\Omega_\mathrm{m}$                                                       & 0.2793              \\
$\Omega_\Lambda$                                                          & 0.7207              \\
$\Omega_\mathrm{b}$                                                       & 0.0463              \\
$h \equiv H_0/(100 \, \mathrm{km}\,\mathrm{s}^{-1} \, \mathrm{Mpc}^{-1})$ & 0.7                 \\
$\sigma_8$                                                                & 0.821              \\
$n_s$                                                                     & 0.972              \\
\hline{}
\end{tabular}
\end{table}

\begin{figure*}
\centering 
\includegraphics[width=0.95\textwidth]{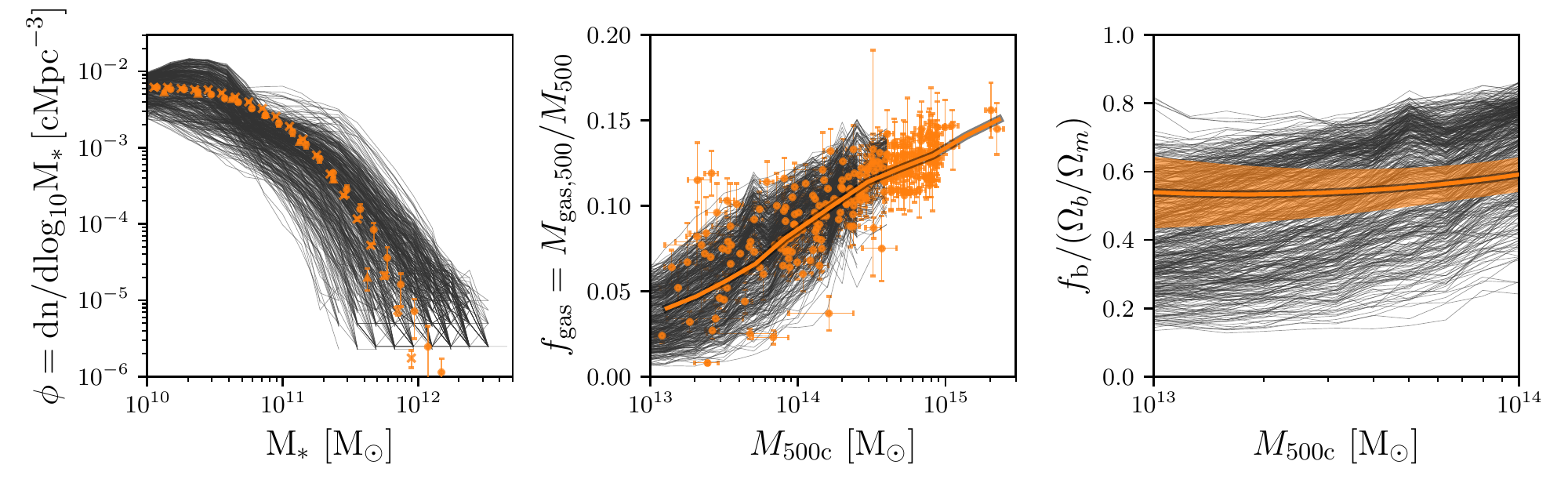}
 \vspace{-1 em}
 \caption{Galaxy stellar mass function \textit{(left)}, median gas fraction \textit{(middle)}, and median total baryon
 fraction \textit{(right)} as a function of halo mass for all 400 simulations at redshift $z=0.125$. Observational data with their associated uncertainties are shown in orange for the stellar mass function \protect\citep{baldry_galaxy_2012,bernardi_massive_2013,moustakas_primus:_2013,2022MNRAS.513..439D}, for the gas fractions of groups and clusters \protect\citep{2006ApJ...640..691V,2008ApJS..174..117M,2009A&A...498..361P,2009MNRAS.399..239R,2009ApJ...693.1142S,2012ApJ...745L...3L,2013ApJ...778...14G,2013MNRAS.429.3288S,2015A&A...573A.118L,2017MNRAS.469.3489P}, the latter of which are derived from resolved X-ray observations, and the fit to the median baryon fraction from the latest HSC-XXL weak gravitational lensing data from \protect\citet{Akino_2022}, including a correction for the contribution of blue galaxies and the diffuse intracluster light. The light shaded region encloses the $1\sigma$ uncertainty. The orange line in the middle panel highlights the median $f_\mathrm{gas}-M_{500\mathrm{c}}$ relation from observations, including a correction from the inferred scaling relations from the X-ray selected sample of galaxy groups and clusters in \protect\citet{2020MNRAS.492.4528S}. Comparing the solid black curves to the solid orange curve (i.e., median relations) in the middle panel, the simulations conservatively bracket the observed gas fractions, the observed galaxy stellar mass function (left panel), as well as the median total baryon fraction (right panel).
 }
 \label{fig:sims}
\end{figure*}

The Boltzmann code \textsc{camb}\footnote{{\href{https://camb.info/}{https://camb.info/}}} \citep[][]{lewis_efficient_2000} was used to compute the transfer functions which were supplied to a modified version of the \textsc{N-GenIC}\footnote{{\href{https://github.com/sbird/S-GenIC}{https://github.com/sbird/S-GenIC}}} code to include second-order Lagrangian Perturbation Theory to create the initial conditions at a starting redshift of $z=127$. When producing initial conditions for hydrodynamical simulations, we use the separate transfer functions computed by \textsc{camb} for each individual component (i.e., baryons and dark matter). Similarly, in order to avoid any offset in the amplitude of the matter power spectrum of the hydro simulations with respect to their dark matter-only counterpart at large scales, for the dark matter-only version of the simulations, we generated two separate fluids, one with the dark matter transfer function and the other with the baryon transfer function \citep{valkenburg_accurate_2017,van_daalen_2020}. Additionally, the same random phases were used to generate each set of initial conditions. Hence, comparisons made between the different simulations are not subject to cosmic variance complications.

The simulation suite was run with a modified version of the \textsc{gadget-}${\scriptstyle 3}$ smoothed particle hydrodynamics (SPH) code (last described by \citealt{springel_cosmological_2005}) and includes a full treatment of gravity and hydrodynamics. Specifically, we use version of \textsc{gadget-}${\scriptstyle 3}$ that was modified for the EAGLE project \citep{schaye_eagle_2015}.

In order to examine the potential effects of different hydrodynamic schemes, the ANTILLES suite comprises a set of simulations that use the standard \textsc{gadget} flavour of SPH, as as well as set that uses the more recent state-of-the-art \textsc{anarchy} formulation. The improvements within \textsc{anarchy} include the use of the pressure-entropy formulation of SPH derived by \cite{hopkins_general_2013}, the artificial viscosity switch from \cite{cullen_inviscid_2010}, an artificial conduction switch similar to that of \cite{price_modelling_2008}, the $\mathcal{C}_2$ kernel of \cite{wendland_piecewise_1995}, and the time-step limiters of \cite{durier_implementation_2012}.

Important non-gravitational processes, such as stellar and AGN feedback, that are not resolved by the simulations are implemented as subgrid physical models. A full description of these subgrid models can be found in \citet{schaye_eagle_2015}. In summary:
\begin{enumerate}
    \item Radiative cooling and photoheating are implemented element-by-element as in \cite{wiersma_effect_2009}, including the 11 elements found to be important, namely, H, He, C, N, O, Ne, Mg, Si, S, Ca, and Fe. Hydrogen reionization is implemented by switching on the full \cite{haardt_modelling_2001} background at redshift $z=7.5$.
    \item Star formation is implemented stochastically following the pressure-dependent Kennicutt-Schmidt relation as in \cite{schaye_relation_2008}. Above a density threshold $n^*_\mathrm{H}=0.1 \, \mathrm{cm}^{-1}$, which is designed to track the transition from a warm atomic to an unresolved cold molecular gas phase \citep{schaye_star_2004}, gas particles have a probability of forming stars determined by their pressure.  We use a single fixed density threshold for these simulations, as opposed to a metallicity-dependent threshold in EAGLE.
    \item Time-dependent stellar mass loss due to winds from massive stars and AGB stars, core collapse supernovae and type Ia supernovae, is tracked following \cite{wiersma_chemical_2009}. 
    \item Stellar feedback is implemented using the kinetic wind model of {}\cite{dalla_vecchia_simulating_2008}, which differs from the thermal implementation used in EAGLE.  This is motivated by resolution considerations, particularly that relative high resolution is required for efficient thermal feedback due to its stochastic implementation.  We instead adopt a kinetic implementation, as also adopted in BAHAMAS.
    \item Seed BHs of mass $\mathrm{M} = 1 \times 10^6 \mathrm{M}_\odot$, are placed in haloes with a mass greater than $2.75 \times 10^{11} \mathrm{M}_\odot$ (corresponding to $\approx 50$ DM particles) and tracked following the methodology of \citet{springel_modelling_2005} and \citet{booth_cosmological_2009}. Once seeded, BHs can grow via Eddington-limited gas accretion, at a rate which is proportional to the Bondi-Hoyle-Lyttleton rate, as well as through mergers with other BHs following \cite{booth_cosmological_2009}.
    \item Feedback from AGN is implemented following the stochastic heating scheme. A fraction of the accreted gas onto the BH is released as thermal energy with a fixed heating temperature into the surrounding gas following \cite{booth_cosmological_2009}.
\end{enumerate}

 In order to explore a wide feedback landscape that brackets current observational constraints (with their associated uncertainties) on the stellar and gas fractions, we systematically vary the main subgrid parameters governing the efficiencies of stellar and AGN feedback. In particular, the galaxy (star) formation efficiency is quite sensitive to variations of the wind velocity ($v_w$) and mass-loading ($\eta_w$) parameters. These parameters have a significant effect on the star formation histories of galaxies and the shape of the GSMF, especially at the low-mass end where stellar feedback is expected to dominate \citep[e.g.][]{schaye_physics_2010,mccarthy_bahamas_2017}. For AGN feedback, the two main parameters that determine how frequent and energetic the feedback events are, are the number of neighbouring gas particles to heat ($n_\mathrm{heat}$) and the temperature increase of such particles ($\Delta T_\mathrm{heat}$). In particular, changes in $\Delta T_\mathrm{heat}$ have a significant impact on the gas mass fraction of groups and clusters of galaxies \citep[][]{mccarthy_bahamas_2017}, which can be understood by recognising that the likelihood of significant gas ejection will be favourable if $\Delta T_\mathrm{heat} \gg T_\mathrm{vir}$. Finally, in the \citet{booth_cosmological_2009} accretion model implemented in the simulations, the ``boost factor'', $\alpha$, relative to pure Bondi-Hoyle accretion is a power-law function of the local density for gas above a pivot point, $n^*_\mathrm{H,BH}$, which is set to the star formation threshold $n^*_\mathrm{H}=0.1 \mathrm{cm}^{-3}$, as the simulations resolve lower densities, where no cold, molecular gas phase, is expected. At higher densities, the simulations lack the cold gas phase, which would boost the Bondi-Hoyle rate due to its sound speed dependence \citep{schaye_star_2004,booth_cosmological_2009}. We set the power-law exponent to $\beta=2$ for high densities, and the boost factor is set to go to unity at low density, where the simulations are well resolved. Nevertheless, for gas at $10^4$ K, at the particle resolution of the simulations, we only resolve the Jeans length for densities $<10^{-3} \mathrm{cm}^{-3}$. Hence, we explore boosting the accretion rate at lower densities to compensate for the lack of resolution. We find that changing the pivot point, $n^*_\mathrm{H,BH}$, can have a significant impact at the mass-scale at which AGN feedback becomes efficient, effectively changing the shape of the knee of the GSMF. The five parameters varied in this study, with their respective range of values, are provided in \cref{tab:params}.  

\begin{table}
\centering
\caption{Subgrid parameters varied in this study.$v_w$ and $\eta_w$ are the wind velocity and mass-loading factor used in the kinetic wind stellar feedback model as per \citet{dalla_vecchia_simulating_2008}. $\Delta T_\mathrm{heat}$ is the temperature increase of gas particles during BH feedback events, and $n_\mathrm{heat}$ is the number of neighbouring gas particles to be heated. $n^*_\mathrm{H,BH}$ is the density threshold above which the BH accretion rate is boosted in the \citet{booth_cosmological_2009} accretion model due to the lack of resolution of the cold gas phase at high densities.}\label{tab:params}
\begin{tabular}{lc}
\hline
Parameter                                                       & Range                     \\ \hline
$v_w$ [km/s]                                                    & $[50, 350]$               \\
$\eta_w$                                                        & $[1, 10]$                 \\
$\mathrm{log}_{10}(\Delta T_\mathrm{heat} [\mathrm{K}])$        & $[7, 8.5]$                \\
$n_\mathrm{heat}$                                               & $[1, 30]$                 \\
$\mathrm{log}_{10}(n^*_\mathrm{H,BH} [\mathrm{cm}^{-3}])$       & $[-3, -1]$                \\
\hline        
\end{tabular}
\end{table}

In order to ensure a full parameter space coverage, we employed a Latin hypercube sampling with multidimensional uniformity (LHSMDU) developed by \citet{DEUTSCH2012763}. We used 200 nodes to uniformly sample the parameter space. Finally, we ``mirrored'' these 200 simulations using the exact same subgrid physics implementation and sampled parameters, but using a standard \textsc{gadget} flavour of SPH, as opposed to the more recent state-of-the-art \textsc{anarchy} formulation. This gives a total of 400 simulations in the ANTILLES sample. 

The GSMF, median gas fraction, and median total baryon fraction as a function of halo mass spanned by the simulations at redshift $z=0.125$ is shown in \cref{fig:sims}. The figure shows how our parameter space exploration easily brackets the current observational constraints. Note that when comparing the simulations to the observations in the middle panel (gas mass fractions), one should compare the solid black curves, which represent the median relations of each of the 400 simulations, to the solid orange curve, which represents the median relation from resolved X-ray observations. (The orange data points represent individual observed groups and clusters, we do not show individual simulated clusters for clarity, but we note the intrinsic scatter in the simulated relations is similar to the observed intrinsic scatter). Thus, the simulations conservatively bracket both the observed gas fractions and the observed stellar mass function. If one adopts the statistical uncertainties on the median baryon mass fractions\footnote{We note that for our simulations, we include all stellar and gas particles within a spherical overdensity radius. Hence, in order to make reasonable comparisons with the fits in \citet{Akino_2022}, we included an additional 15\% contribution to the total stellar masses from the contribution of blue galaxies, and 30\% additional stellar mass to the brightest cluster galaxies (BCGs) to account for the diffuse intracluster light (ICL, see \citealt{Akino_2022}).} from the recent study of \citet{Akino_2022}, then our simulations span a range of approximately $-7\sigma$ to $+6\sigma$ with respect to the observed baryon fraction at a mass scale of $\approx10^{14}$ M$_\odot$.

Note that our desire to span a much wider range of baryon fractions than is apparently allowed by current observations is motivated by two factors: i) the low-redshift data we compare to may have non-negligible biases; and ii) the baryon fractions of higher redshift (e.g., $z \approx 0.5-1$) groups and clusters, which give rise to much of the lensing signal, is not well constrained at present by observations, thus we want a range of simulated behaviours that is wide enough to hopefully encapsulate future measurements of high-redshift systems as well. With regards to possible biases in current data, observational measurements are always subject to both random and systematic errors. For instance, the observationally-derived stellar masses are subject to systematic errors originating from, e.g., stellar population modelling, spectral energy distribution fitting, surface brightness profile fitting and corrections for dust extinction.  Gas fraction measurements from X-ray data of groups and clusters are subject to uncertainties in, e.g., deviations from spherical symmetry and hydrostatic equilibrium, and modelling of selection effects, particularly in the group regime.

Finally, we note that the prior ranges on each of the subgrid parameters were essentially selected by examining the baryon fraction results from a small number of test simulations, so it is no surprise that the resulting hypercube spans a large range of baryon fractions.

Note also that while we use the combined set of 400 simulations to construct our model for the suppression of the matter power spectrum (below), we have also explored analysing the two SPH suites separately. While for a given set of subgrid parameters the choice of SPH flavour can affect the resulting baryon fractions, the reaction in terms of $P(k)$ is independent of the choice of SPH flavour at a given baryon fraction. This is consistent with the findings of \citet{van_daalen_2020} and allows us to construct a single (more accurate) model in terms of baryon fractions using the entire set of 400 ANTILLES simulations.

\begin{figure}
\centering 
\includegraphics[width=0.48\textwidth]{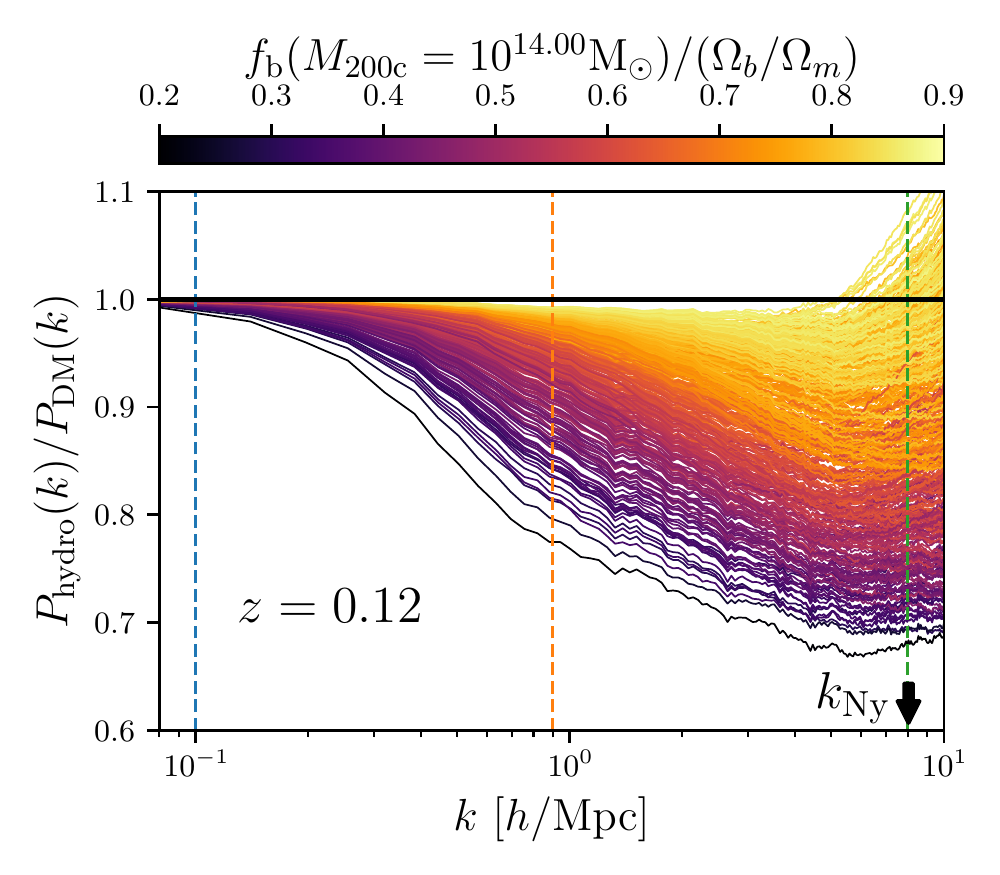}
 \vspace{-1.5em}
 \caption{Fractional impact of baryons on the total matter power spectrum for the 400 models at redshift $z=0.125$. Colour-coding represents the total baryon fraction of haloes of mass $M_{200\mathrm{c}} = 10^{14} \mathrm{M}_\odot$. Vertical dashed lines indicate 3 equally log-spaced scales, $k=0.1, 0.9$ and $8.0$ ${h \,\mathrm{Mpc}^{-1}}$, shown in \cref{fig:scatter}. The black arrow indicates the Nyquist frequency of the simulations $k_{\mathrm{Ny}}$.}
 \label{fig:PS}
\end{figure}

\begin{figure*}
\centering 
\includegraphics[width=0.95\textwidth]{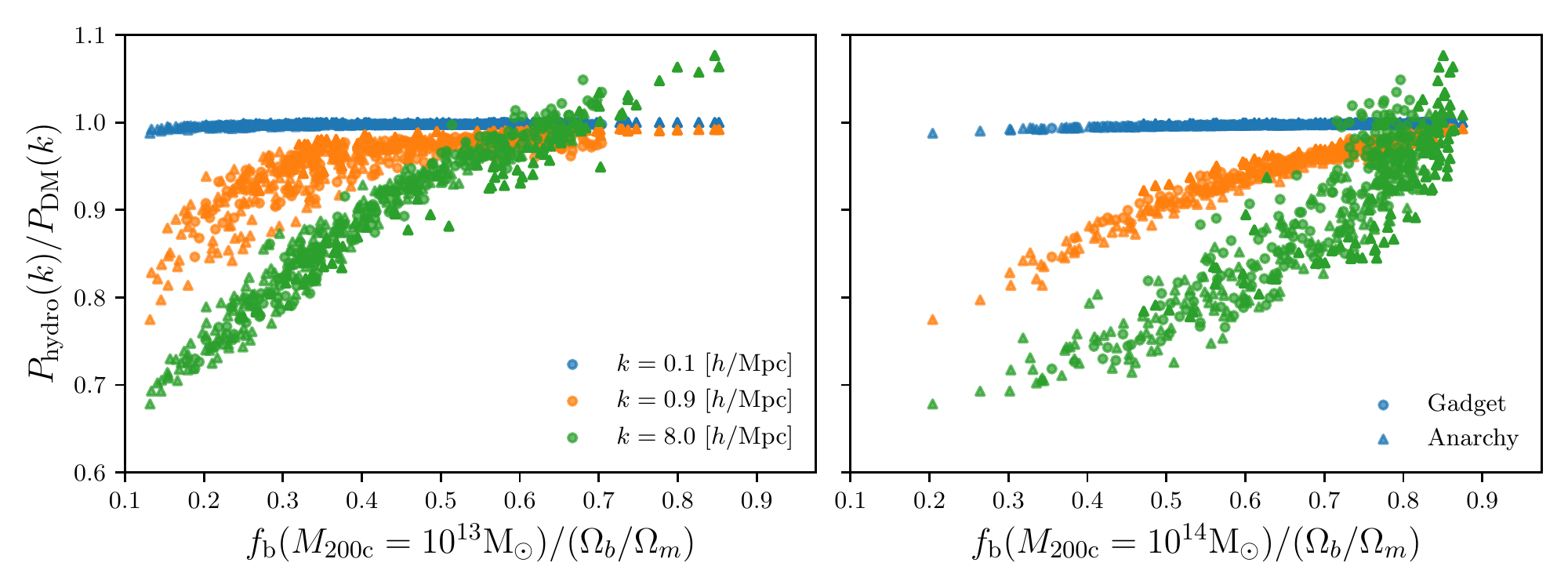}
 \vspace{-1 em}
 \caption{Fractional impact of baryons on the total matter power spectrum at $k=0.1, 0.9$ and $8.0$ ${h \,\mathrm{Mpc}^{-1}}$ as a function the total baryon fraction of haloes of mass $M_{200\mathrm{c}} = 10^{13} \mathrm{M}_\odot$ \textit{(left)} and $M_{200\mathrm{c}} = 10^{14} \mathrm{M}_\odot$ \textit{(right)}. Colours indicate the suppression at the 3 equally spaced scales in logarithmic scale ($\log_{10}k$) in \cref{fig:PS}. Each symbol represents one of the 400 models. The two sets of SPH schemes used for the simulations are shown with different symbols. The suppression of the power spectrum at a given scale correlates better with the total baryon fraction of haloes of different mass. For instance, the suppression at $k=0.9$ ${{h \,\mathrm{Mpc}^{-1}}}$ \textit{(orange symbols)} is better correlated with the baryon fraction of haloes of mass $M_{200\mathrm{c}} = 10^{14} \mathrm{M}_\odot$ \textit{(right)} than with $M_{200\mathrm{c}} = 10^{13} \mathrm{M}_\odot$ \textit{(left)}. The converse is true for $k=8.0$ ${h \,\mathrm{Mpc}^{-1}}$ \textit{(green symbols)}, while the strength of the correlation seems insensitive to the two masses shown for $k=0.1$ ${h \,\mathrm{Mpc}^{-1}}$ \textit{(blue symbols)}.}
 \label{fig:scatter}
\end{figure*}

\section{Modelling baryon physics effects on $P(k)$}\label{sec:model}

We begin by analysing the relative (fractional) impact of baryon physics on the total matter power spectrum for the 400 simulations. \Cref{fig:PS} shows the ratio of power spectra\footnote{We characterise the ratio in terms of the power spectrum, $P(k)$. But note that since $P(k) \propto \Delta^2(k)$, our results are equivalent to a ratio of $\Delta^2_\mathrm{hydro}(k)/\Delta^2_\mathrm{DM}(k)$.} from the different simulations with respect to a DM-only counterpart at redshift $z\approx0.1$. The wide range of variations in the feedback models in our simulations gives rise to a large diversity of impacts on $P(k)$, in agreement with previous studies that show that the effect of baryons on the power spectrum depends strongly on the adopted baryon physics \citep[e.g.][]{van_daalen_effects_2011,mummery_separate_2017,springel_first_2018,chisari_impact_2018,chisari_modelling_2019,van_daalen_2020}. Given the selection of prior ranges for each subgrid parameter to conservatively bracket existing observational constraints and their corresponding uncertainties, our simulation suite encompasses baryonic effects that surpass observational bounds, with a range of parameters more extreme than those used in most cosmological simulations (see e.g. \citealt{troster_2022}). The effect of AGN feedback is of particular importance, due to its ejective nature at early times \citep{mccarthy_gas_2011}. The removal of large quantities of gas not only affects the distribution of the remaining gas, but it also has a gravitational effect on the dark matter, causing it to expand as well.  This effect is referred to as the `back reaction' on baryons on the dark matter (e.g., \citealt{van_daalen_effects_2011}).  The net result of baryon ejection and dark matter expansion is that $P(k)$ can be suppressed by a non-negligible amount (relative to forthcoming statistical errors from cosmic shear surveys) out to wavenumbers of $k \sim 0.1$ $h$ Mpc$^{-1}$, corresponding to physical scales of $\sim 30$ Mpc/$h$.

In the following sections we will present our model based on the correlation between the power spectrum suppression and the total baryon fraction of haloes of different masses. Note that in this study, halo masses are measured within both $R_{200\mathrm{c}}$ and $R_{500\mathrm{c}}$, i.e. the radius within which the mean density is 200 or 500 times the critical density of the Universe respectively. While both spherical overdensity apertures provide similar results, we find that using $R_{200\mathrm{c}}$ gives slightly better accuracy. Hence, in the following sections, we will presents all of our results and plots in terms of $R_{200\mathrm{c}}$, while also provide the best fitting parameters for our model in both $R_{200\mathrm{c}}$ and $R_{500\mathrm{c}}$.

\begin{figure*}
\centering 
\includegraphics[width=0.95\textwidth]{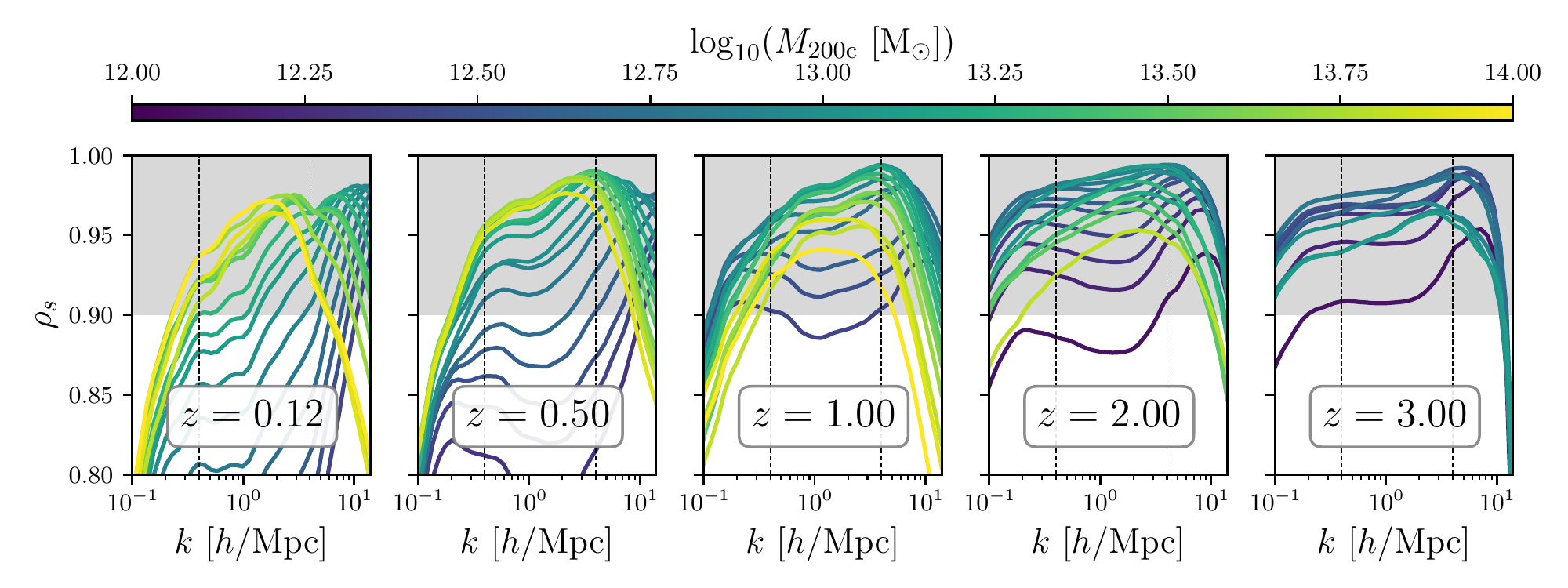}
 \vspace{-1 em}
 \caption{Spearman's rank correlation coefficient $\rho_{s}$, between the suppression of the total matter power spectrum and the total baryon fraction of haloes of different mass as a function of scale $k$. The five panels represent different redshifts and colour-coding represents different halo masses $M_{200\mathrm{c}}$. The grey shaded area indicates where $\rho_{s} \geq 0.90$. Independent of redshift, for small scales ($k \gtrsim 4 \, {h \,\mathrm{Mpc}^{-1}}$), the baryon fraction of progressively lower mass haloes provide better correlation for increasing $k$. For intermediate scales ($0.4 \lesssim k [{h \,\mathrm{Mpc}^{-1}}] \lesssim4$), haloes with a characteristic mass (which evolves with redshift) provide the best correlation. For $k \lesssim 0.4 \, {h \,\mathrm{Mpc}^{-1}}$, the strength of the correlation decreases rapidly regardless of the halo mass.}
 \label{fig:spearman}
\end{figure*}

\subsection{The correlation between the power spectrum suppression and the total baryon fraction}

Recently, \citet{van_daalen_2020} presented an empirical model that uses the mean baryon fraction of group mass haloes ($M_{200\mathrm{c}} = 10^{14} \mathrm{M}_{\odot}$) as a predictor for the power suppression for scales $k \lesssim 1.0 \,h$ Mpc$^{-1}$, achieving an accuracy of $\approx 1\%$.  Fundamentally, the effect of baryon physics on the matter power spectrum is a gravitational process, which is why we expect the baryon fraction of haloes (which is a direct measure of the mass that has been removed) that contribute most significantly to $P(k)$ to be a strong predictor of the effects on $P(k)$.

Taking our cue from \citet{van_daalen_2020}, in \cref{fig:PS} we colour code each run by its median total baryon fraction of haloes of mass $M_{200\mathrm{c}} = 10^{14} \ \mathrm{M}_{\odot}$, and normalised by the universal baryon fraction ($\Omega_b/\Omega_m$). In agreement with \cite{van_daalen_2020}, the figure shows a clear gradient of the power spectrum suppression as a function of the total baryon fraction of group mass of haloes.  At closer inspection, the correlation is not perfect at small scales of $k \geq 3 \,h$ Mpc$^{-1}$.  Nevertheless, we will show below that a simple and accurate model for the $P(k)$ suppression can be formulated in terms of the median baryon fraction. 

In this work, we aim to extend the \cite{van_daalen_2020} model to smaller scales and higher redshifts, which we will achieve by taking into account the increasing importance of lower mass haloes as we push in both directions.  In order to assess the strength of the correlation between the baryon fraction of haloes of different mass, we compute the power spectra at three equally spaced scales in $\mathrm{log}_{10}k$ ($k=0.1, 0.9$ and $8.0$ $h$ Mpc$^{-1}$), which are shown as vertical dashed lines in \cref{fig:PS}. We show in \cref{fig:scatter} the fractional impact of baryons on the total matter power spectrum as a function the total baryon fraction for two different halo of masses ($10^{13} \mathrm{M}_\odot$ and $10^{14} \mathrm{M}_\odot$) for these three scales. The colours correspond to the same scales of the vertical lines in \cref{fig:PS}. The two sets of SPH schemes used for the simulations are shown with different symbols.

\cref{fig:scatter} shows that the suppression of the power spectrum at a given scale correlates better with the total baryon fraction of haloes of different mass. For instance, the suppression at $k=0.9 h$ Mpc$^{-1}$ \textit{(orange symbols)} is better correlated with the baryon fraction of haloes of mass $10^{14} \mathrm{M}_\odot$ \textit{(right)} than with $10^{13} \mathrm{M}_\odot$ \textit{(left)}. On the other hand, for small scales, $k=8.0 h$ Mpc$^{-1}$ \textit{(green symbols)}, the suppression is better correlated with the baryon fraction of lower mass haloes \textit{(left)}. This dependency can be explained in terms of the differing relative contribution of haloes of different mass to the total matter power spectrum (e.g., \citealt{van_daalen_contributions_2015,mead_hydrodynamical_2020}). The strength of the correlation for very large scales, $k=0.1$ ${h \,\mathrm{Mpc}^{-1}}$ \textit{(blue symbols)}, seems insensitive to choice of halo mass (for the two masses shown here) which is mainly just because feedback does not significantly alter the power spectrum on these scales, so there is little dynamic range in $P_\mathrm{hydro}(k)/P_\mathrm{DM}(k)$ to couple to the baryon fractions.

Note that whilst the selection of SPH flavor for a particular set of subgrid parameters may have an impact on the resulting baryon fractions, the response exhibited in $P(k)$ is invariant with regard to the selection of SPH flavor at a given baryon fraction. This is consistent with the findings of \citet{van_daalen_2020}.

In \cref{fig:spearman} we quantify the strength of the correlation between the suppression of the total matter power spectrum and the total baryon fraction of haloes of different mass as a function of scale $k$ using the Spearman's rank correlation coefficient, $\rho_s$. We calculate the correlation coefficient at 40 equally spaced scales\footnote{We re-bin the power spectra and `smooth' them by calculating the median power for each bin.} in $\log_{10} k$ in the range $0.1 \leq k \,\, [h\,\mathrm{Mpc}^{-1}] \leq 10$. Each panel represents a different redshift in the simulations. 

For each redshift, the lines are colour coded by the halo mass used to estimate the baryon fraction.  We use a mass bin width of 0.1 dex around each indicated halo mass.  To help guide the eye, we have highlighted the area where $\rho_{s} \geq 0.90$, i.e. where the correlation between the suppression of the total matter power spectrum and the total baryon fraction is exceptionally strong, and it can be well described using a monotonic function. 

To guide the eye, we have added three vertical lines in \cref{fig:spearman} to roughly divide small, intermediate, and large scales:

\begin{enumerate}
    \item \textit{Small scales} ($k \gtrsim 4 \, h\,\mathrm{Mpc}^{-1}$): the baryon fraction of progressively lower mass haloes provides a better correlation with the suppression for increasing $k$. The converse is also true; i.e., the strength of the correlation using higher mass haloes decreases for increasing $k$. 
    \item \textit{Intermediate scales} ($0.4 \lesssim k \,\, [h\,\mathrm{Mpc}^{-1}] \lesssim 4$): haloes with a characteristic mass (which evolves with redshift) provide the best correlation. 
    \item \textit{Large scales} ($k \lesssim 0.4 \, h\,\mathrm{Mpc}^{-1}$): the strength of the correlation decreases rapidly regardless of the halo mass. This is expected, as for (very)large scales, the suppression of the total matter power spectrum should be close to unity independent of the baryon fraction of haloes (non-monotonic, see \cref{fig:scatter}).
\end{enumerate}

\begin{figure}
\centering 
\includegraphics[width=0.48\textwidth]{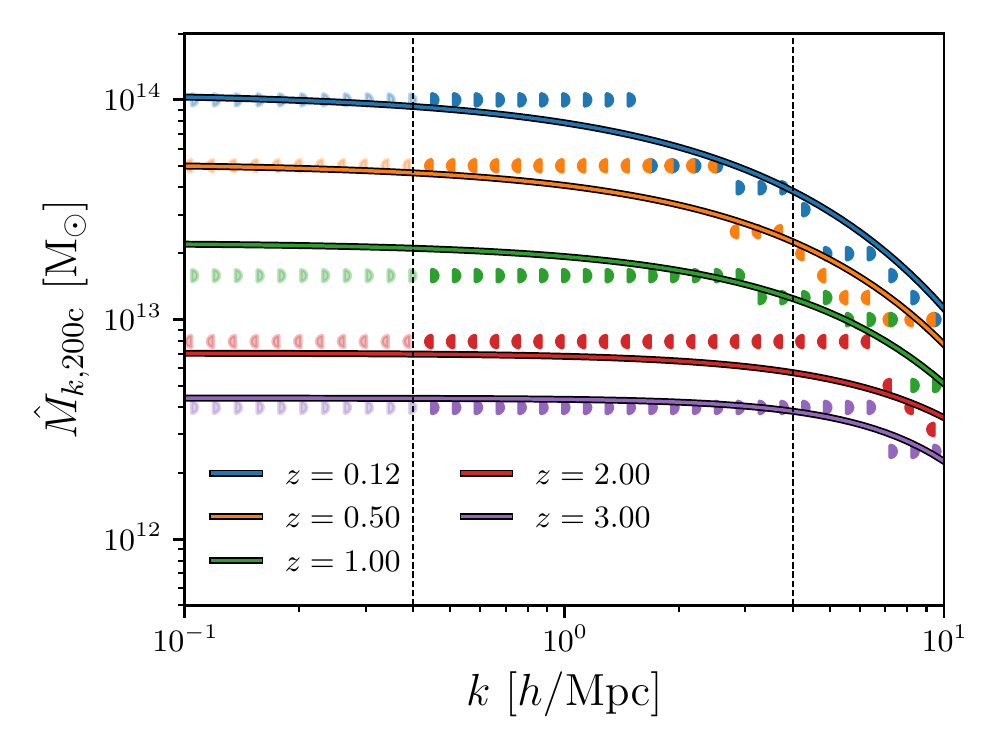}
 \vspace{-1.5em}
 \caption{Optimal mass $\hat{M}_{k,200\mathrm{c}}$ as a function of scale $k$ at five different redshifts. Coloured symbols represent the halo mass that provides the best correlation between the power spectrum suppression and the baryon fraction of such haloes at each $k$ bin. Solid lines show the best fit using \cref{eq:optimal_mass_fit}. For small scales ($k \gtrsim 4 \, {h \,\mathrm{Mpc}^{-1}}$), the optimal mass decreases for increasing $k$. For intermediate scales ($0.4 \lesssim k \,\, [h\,\mathrm{Mpc}^{-1}] \lesssim 4$), haloes with a characteristic mass provide the best correlation (i.e., the optimal mass plateaus). For large scales the suppression of the total matter power spectrum should approach unity, independent of the baryon fraction of haloes (see \cref{fig:scatter,fig:spearman}). Therefore, in order to avoid noise while fitting, we assume that $\hat{M}_{k}$ remains constant for $k \leq 0.4 \, {h \,\mathrm{Mpc}^{-1}}$ and data points are show with faint colours.}
 \label{fig:best_mass}
\end{figure}

\begin{figure}
\centering 
\includegraphics[width=0.48\textwidth]{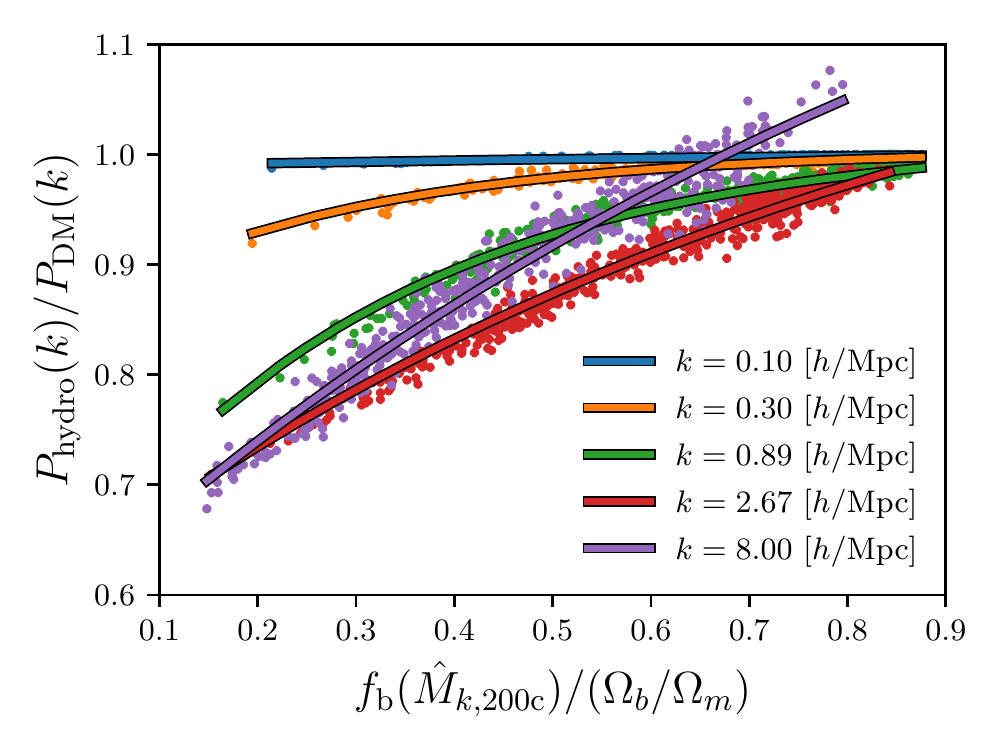}
 \vspace{-1.5em}
 \caption{Suppression of the total matter power spectrum at redshift $z=0.125$ as a function the total baryon fraction using the optimal mass of haloes $\hat{M}_{k,200\mathrm{c}}$ for each corresponding scale $k$. Colours indicate the suppression at 5 equally spaced scales in logarithmic scale ($\log_{10}k$). Each symbol represents one of the 400 models. Solid lines show the best fit using \cref{eq:sup_fit} without a scale and redshift dependence, i.e. each scale shown has been fitted using a simple exponential plateau function.}
 \label{fig:fit}
\end{figure}

\begin{figure*}
\centering 
\includegraphics[width=0.99\textwidth]{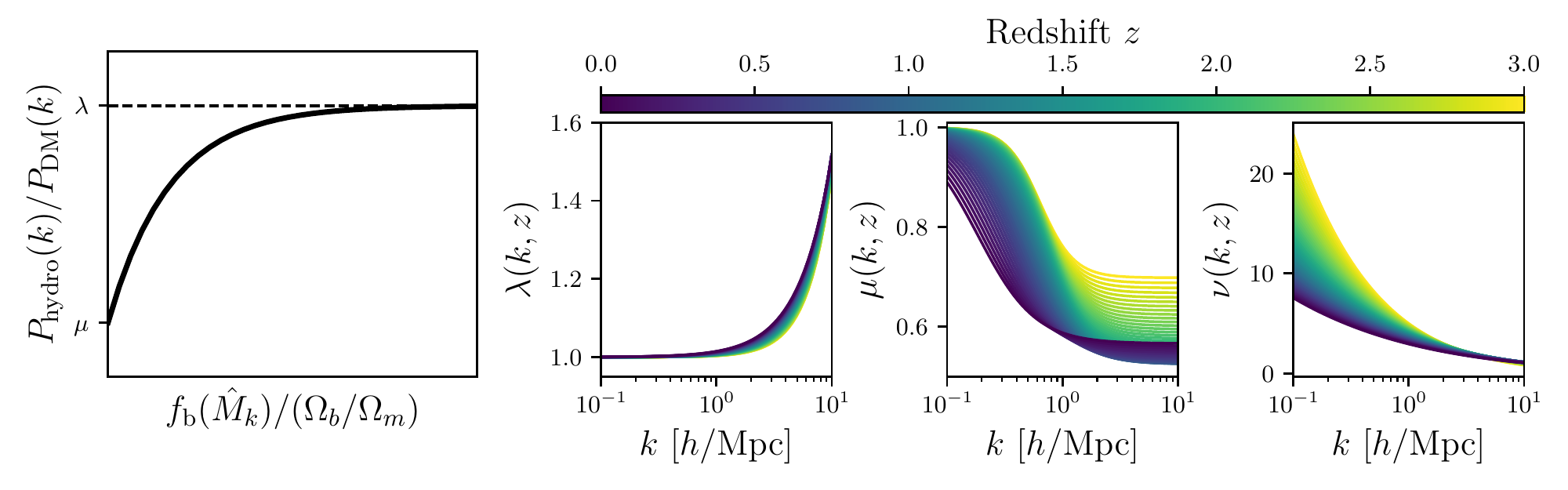}
 \vspace{-1 em}
 \caption{\textit{Left:} Illustration of the suppression of the total matter power spectrum as a function of $\tilde{f}_b$ as parametrised by \cref{eq:sup_fit}. \textit{Right:} Scale and redshift dependence of the best fit parameters in \cref{tab:sup_fit}. $\lambda(k,z)$ is the asymptote of the function in \cref{eq:sup_fit} for large $\tilde{f}_b$ values, modelled using an exponential functional form \pcref{eq:func_forms1}, $\mu(k,z)$ is the suppression for $\tilde{f}_b = 0$, modelled using a logistic functional form \pcref{eq:func_forms2}, and $\nu(k,z)$ is the rate of change, modelled with log-normal functional form \pcref{eq:func_forms3}.}
\label{fig:model_params}
\end{figure*}

\subsection{The optimal halo mass}

As we aim to determine the halo masses that best predict the suppression of the power spectrum at each scale, we define the \textit{optimal mass}, $\hat{M}_{k,200\mathrm{c}}$, as the halo mass used to measure the baryon fraction that maximises $\rho_s$ as a function of scale and redshift:
\begin{equation}\label{eq:optimal_mass}
    \hat{M}_{k,200\mathrm{c}}(k,z) = M_{200\mathrm{c}}\big\rvert_{\max \{\rho_s(k,z)\}} \ \ .
\end{equation}

\Cref{fig:best_mass} shows $\hat{M}_{k,200\mathrm{c}}$ as a function of scale $k$ for the five redshifts shown in \cref{fig:spearman}. We used alternating filled semicircles to allow us to show overlapping points from different redshift bins. For each redshift, the shape of the relation roughly reflects the three distinct behaviours described above: at small scales ($k \gtrsim 4 \, {h \,\mathrm{Mpc}^{-1}}$), the baryon fraction of haloes decreasing in mass provide a better correlation for increasing $k$; at intermediate scales ($0.4 \lesssim k \,\, [h\,\mathrm{Mpc}^{-1}] \lesssim 4$), haloes with a characteristic mass provide the best correlation (i.e., the optimal mass plateaus); and at large scales, the strength of the correlation decreases rapidly regardless of the halo mass, as the total matter power spectrum suppression is close to unity, independent of the baryon fraction of haloes (see \cref{fig:scatter,fig:spearman}). In order to avoid noise while fitting, without loss of generality, we assume that $\hat{M}_{k}$ remains constant for $k \leq 0.4 \, {h \,\mathrm{Mpc}^{-1}}$ (data points are show in faint colours). Note that this is also consistent with the results of \cite{van_daalen_2020}, i.e., to a good approximation, the total baryon fraction for a single halo mass provides enough information to predict the power spectrum suppression due to baryons for $k \lesssim 1.0 \,{h \,\mathrm{Mpc}^{-1}}$. 

It is worth noting that the optimal mass exhibits a monotonic behaviour as function of redshift. Therefore, we approximate $\hat{M}_{k,200\mathrm{c}}$ using the following exponential plateau functional form, 
\begin{equation}\label{eq:optimal_mass_fit}
    \log_{10}\qty(\hat{M}_{k,200\mathrm{c}}(k,z)) = \alpha(z)  - \qty[\alpha(z) - \beta(z)] k^{\gamma(z)} \ \ ,
\end{equation}
where $\alpha(z)$ is the asymptote of the function as ${k {\to} 0}$, $\beta(z)$ is the value for $k=1$, and $\gamma(z)$ characterises the rate of change. We impose a simple polynomial ansatz for the redshift dependence of these parameters as follows,
\begin{align}\label{eq:z_dep}
\begin{split}
    X(z) = \sum_{i=0}^2{X_{i}(1+z)^{i}} \ \ ,
\end{split}
\end{align}
where $X=\{\alpha, \beta, \gamma\}$.

We use a simple least squares estimator to fit the simulation data and we provide the best-fitting parameter values in \cref{tab:opt_mass}. The solid lines in \cref{fig:best_mass} show the best fit using \cref{eq:optimal_mass_fit}. 

\begin{table}
\centering
\caption{Best-fit parameters for the optimal mass in \cref{eq:optimal_mass_fit}, expressed either in terms of $M_{200c}$ or $M_{500c}$. For the exponential plateau functional form, $\alpha$ is the asymptote of the function as ${k {\to} 0}$, $\beta$ is the value for $k=1$, and $\gamma$ characterises the rate of change. The redshift dependence of the parameter is given in \cref{eq:z_dep}.}
\label{tab:opt_mass}
\begin{tabular}{cllll}
\hline
\textbf{Spherical overdensity}          & &\multicolumn{3}{c}{\textbf{Parameter}}    \\ %\hline
                                        & \textbf{}             & $i=0$ & $i=1$ & $i=2$ \\ \hline
\multirow{3}{*}{\textbf{$M_{200c}$}}    & \textbf{$\alpha_i$}   & 15.243 & -1.243 & 0.148    \\
                                        & \textbf{$\beta_i$}    & 14.969 & -1.099 & 0.129    \\
                                        & \textbf{$\gamma_i$}   & 0.800  & -0.017 & 0.061    \\ \hline
\multirow{3}{*}{\textbf{$M_{500c}$}}    & \textbf{$\alpha_i$}   & 14.783 & -0.999 & 0.120   \\
                                        & \textbf{$\beta_i$}    & 14.620 & -0.913 & 0.108   \\
                                        & \textbf{$\gamma_i$}   & 0.967  & -0.031 & 0.026    
\end{tabular}
\end{table}

In \cref{fig:fit} we use the optimal mass $\hat{M}_{k,200\mathrm{c}}(k,z)$ from \cref{eq:optimal_mass_fit} to compute the suppression of the total matter power spectrum as a function the total baryon fraction at redshift $z=0.125$. Note that we do not assume any functional form for the $f_b - M_\mathrm{halo}$ relation. Rather, we use a piecewise cubic Hermite interpolating polynomial using the narrow binned data from the simulations to compute the baryon fractions at the required optimal mass. 

The figure shows that using $\hat{M}_{k,200\mathrm{c}}(k,z)$ significantly reduces the scatter of the relationship at all scales shown, from $k = 0.1 \, {h \,\mathrm{Mpc}^{-1}}$, up to the one-dimensional Nyquist frequency of the simulations, $k_{\mathrm{Ny}} = \pi N/L \approx 8 \, {h \,\mathrm{Mpc}^{-1}}$, where $N$ is the cube root of the total number of particles, and $L$ is the length of the cubic box. Given the wide range of physical models and the different hydrodynamic schemes used in this study, the fact that such a tight correlation is achieved solely based on the baryon fraction of haloes of different masses is remarkable. One might have expected that a thorough characterisation of the mass density \textit{profiles} around haloes (as opposed to a simple aperture baryon fraction) would have been required to obtain this level of precision. For example, \citet{debackere_impact_2020} have shown that the behaviour of the profiles between $r_{500}$ and $r_{200}$ can affect the matter power spectrum if the profiles are allowed to vary significantly over this range. Evidently, there is a very strong physical correlation between the integrated baryon fractions and the shape of the profile over this range in our simulations, such that most of the physical effect on the power spectrum is encoded in the integrated baryon fractions alone.

\subsection{SP(k) - A model for the suppression of $P(k)$}

We find that, similar to \cref{eq:optimal_mass_fit}, an exponential plateau functional form provides a good approximation to the fractional impact of baryons on the total matter power spectrum:
\begin{equation}\label{eq:sup_fit}
    P_\mathrm{hydro}(k)/P_\mathrm{DM}(k) = \lambda(k,z) - \qty[\lambda(k,z) - \mu(k,z)] \exp[{-\nu(k,z)\tilde{f}_b}] \ \ ,
\end{equation}
where $\tilde{f}_b$ is the baryon fraction at the optimal halo mass normalised by the universal baryon fraction, i.e.: 
\begin{equation}\label{eq:opt_fb}
    {\tilde{f}_b = f_b(\hat{M}_{k,200\mathrm{c}}(k,z))/\qty(\Omega_b/\Omega_m)} \ \ ,
\end{equation}
 and where $\lambda(k,z)$ is the asymptote of the function for large $\tilde{f}_b$ values, $\mu(k,z)$ is the suppression for $\tilde{f}_b = 0$, and $\nu(k,z)$ is the rate of change parameter. In the left panel of \cref{fig:model_params}, we show an illustration of the suppression of the total matter power spectrum as a function of $\tilde{f}_b$, as parametrised by \cref{eq:sup_fit}.  

We use this functional form to fit the 400 simulations at each scale individually. The best fit for each scale at $z=0.125$ is shown as solid lines in \cref{fig:fit}. \Cref{eq:sup_fit}, without a scale and redshift dependence, provides a good approximation to the full range of scales up to the Nyquist frequency. Furthermore, we approximate each of the parameters in \cref{eq:sup_fit} based on their behaviour as a function of scale using an exponential, logistic and log-normal functional forms (respectively):
\begin{align}
    \lambda(k, z) &= 1 + \lambda_a(z) \exp(\lambda_b(z)  \log_{10}(k)) \ \ , \label{eq:func_forms1}\\
    \mu(k, z) &= \mu_a(z) + \frac{1 - \mu_a(z)}{1 + \exp({\mu_b(z) \log_{10}(k) + \mu_c(z)})} \ \ ,\label{eq:func_forms2}\\
    \nu(k, z) &= \nu_a(z) \exp(-{\frac{\qty(\log_{10}(k) - \nu_b(z))^2}{2\nu_c(z)^2}}) \label{eq:func_forms3} \ \ .
\end{align}

We model the evolution of each parameter as a polynomial function in redshift, using \cref{eq:z_dep}, with ${X=\{\lambda_a, \lambda_b, \mu_a, \mu_b, \mu_c, \nu_a, \nu_b, \nu_b\}}$ accordingly. 

In order to reduce the sensitivity to outliers in our wide range of baryon fractions in the 400 simulations, for our parameter estimation utilised a Huber loss function defined as
\begin{equation}\label{eq:huber}
    L_\delta(y,\hat{y}) = 
    \begin{cases}
        \frac{1}{2} (y - \hat{y})^2                                         & \text{for $\left| y - \hat{y} \right| \leq \delta$}\\
        \delta\left(\left| y - \hat{y} \right| - \frac{1}{2}\delta\right)   & \text{otherwise  ,}
       \end{cases}
\end{equation}
where $y_i$ is the measured variable $P_\mathrm{hydro}(k)/P_\mathrm{DM}(k)$ and $\hat{y}_i$ is the predicted values from our model using using \crefrange{eq:sup_fit}{eq:func_forms3}. The value of $\delta$ was chosen to represent the median absolute deviation. We minimised the loss function \cref{eq:huber} at five anchor redshifts ($z=0.125,0.5, 1.0$, $2.0$ and $3.0$). The best-fit parameters are provided in \cref{tab:sup_fit}.

We present a publicly available Python implementation of our model as \texttt{py-SP(k)}. All relevant information about installation and usage can be found at \href{https://github.com/jemme07/pyspk}{https://github.com/jemme07/pyspk}. 

\begin{table}
\centering
\caption{Best-fit parameters for power spectrum suppression in \cref{eq:sup_fit}. For the exponential plateau functional form, $\lambda$ is the asymptote of the function for large $k$ values, $\mu$ is the suppression for $\tilde{f}_b = 0$, and $\nu$ characterises the rate of change. The $k$ scale and redshift dependence of the parameter are given in \cref{eq:z_dep} and \crefrange{eq:func_forms1}{eq:func_forms3}.}
\label{tab:sup_fit}
\begin{tabular}{cllll}
\hline
\textbf{Spherical overdensity}          & &\multicolumn{3}{c}{\textbf{Parameter}}    \\ %\hline
                                        & \textbf{}             & $i=0$ & $i=1$ & $i=2$ \\ \hline
\multirow{8}{*}{\textbf{$M_{200c}$}}    & \textbf{$\lambda_{a,i}$}      & 0.021 & -0.007 & 0.000   \\
                                        & \textbf{$\lambda_{b,i}$}      & 3.087 & 0.452 & 0.001   \\
                                        & \textbf{$\mu_{a,i}$}          & 0.693 & -0.169 & 0.042   \\
                                        & \textbf{$\mu_{b,i}$}          & 3.161 & 0.861 & 0.011   \\
                                        & \textbf{$\mu_{c,i}$}          & 5.532 & -3.086 & 0.508   \\
                                        & \textbf{$\nu_{a,i}$}          & 413.009 & 311.639 & 37.891  \\
                                        & \textbf{$\nu_{b,i}$}          & -11.243 & -0.344 & 0.334   \\
                                        & \textbf{$\nu_{c,i}$}          & 3.476 & -0.018 & -0.082  \\ \hline
\multirow{8}{*}{\textbf{$M_{500c}$}}    & \textbf{$\lambda_{a,i}$}      & 0.019 & -0.007 & 0.000  \\
                                        & \textbf{$\lambda_{b,i}$}      & 2.956 & 0.620 & -0.001  \\
                                        & \textbf{$\mu_{a,i}$}          & 0.715 & -0.192 & 0.049  \\
                                        & \textbf{$\mu_{b,i}$}          & 3.385 & 0.965 & -0.068 \\
                                        & \textbf{$\mu_{c,i}$}          & 4.457 & -2.191 & 0.454  \\
                                        & \textbf{$\nu_{a,i}$}          & 478.864 & 429.887 & 249.256 \\
                                        & \textbf{$\nu_{b,i}$}          & -11.227 & -0.558 & 0.448  \\
                                        & \textbf{$\nu_{c,i}$}          & 3.499 & -0.084 & -0.092 \\
\end{tabular}
\end{table}

The choice of functional form in \cref{eq:sup_fit} provides insight into the effect of baryons on the total matter power spectrum. In the right-hand side of \cref{fig:model_params} we show the scale and redshift dependence of the best fit parameters in \cref{tab:sup_fit}. The figure shows that for very large scales ($k \sim 0.1 \, {h \,\mathrm{Mpc}^{-1}}$), both $\lambda$ and $\mu$ are close to unity, especially at high redshift, i.e. the suppression of the matter power spectrum is a flat function of the baryon fraction of haloes. In other words, at very large scales, baryons do not significantly affect the matter power spectrum. As we examine progressively smaller scales, on the one hand, $\lambda$ is an increasing function of $k$, i.e. models with large median baryon fraction $(\tilde{f}_b \to 1)$ exhibit weaker suppression values, or even a power spectrum enhancement for small scales ($k \gtrsim 4 \, h\,\mathrm{Mpc}^{-1}$). On the other hand, $\mu$ is a decreasing function of $k$, i.e. models with low median baryon fraction $(\tilde{f}_b \to 0)$ present stronger suppression, especially for scales $k \gtrsim 1 \, h\,\mathrm{Mpc}^{-1}$. Finally, the rate of change parameter, $\nu$, dictates how quickly the suppression changes from the value of $\mu$ at low baryon fractions, to $\lambda$ at large baryon fractions. For very large scales, the value of $\nu$ is irrelevant, as both $\lambda$ and $\mu$ are close to unity. For scales $0.2 \lesssim k \,\, [h\,\mathrm{Mpc}^{-1}] \lesssim 1$, $\nu \gg 1$, i.e. there is a rapid transition from strong suppression values at low baryon fractions, to weak suppression values at large baryon fractions. This is indeed the case if we examine the behaviour the suppression for that scales range in \cref{fig:fit} (orange and green lines). There is a rapid transition giving rise to a ``curved'' relationship that asymptotes to the value of $\lambda$ at high baryon fractions. For smaller scales ($k \gtrsim 1 \, h\,\mathrm{Mpc}^{-1}$), {$\nu {\to} 0$}, i.e. there is a slow transition from strong suppression values at low baryon fractions, to weak suppression (or enhancement) values at large baryon fractions. Examining \cref{fig:fit} again, this slow transition gives rise to a more linear relation (rather than curved) for small scales (red and purple lines). 

\begin{figure}
\centering 
\includegraphics[width=0.48\textwidth]{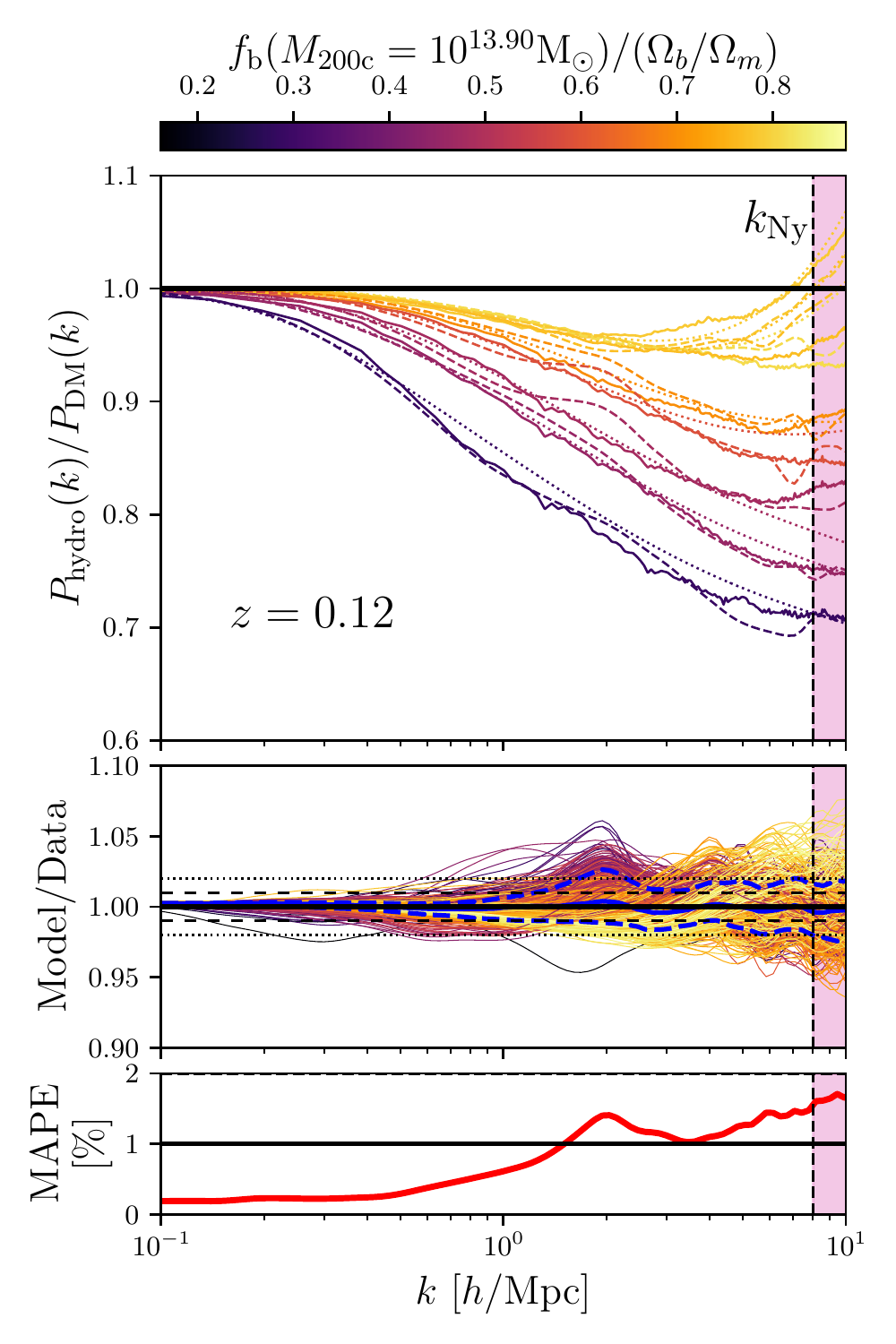}
 \vspace{-1.5em}
 \caption{\textit{Top:} Fractional impact of baryons on the total matter power spectrum for 8 randomly selected samples from the 400 models. Solid lines use power spectra computed from the simulations. Dashed lines represent the best-fitting model obtained with \texttt{SP(k)} using binned data for the $f_b-M_\mathrm{halo}$ relation. Dotted lines use a simple power-law fit to the total baryon fraction instead. Colour-coding represents the total baryon fraction of haloes of mass $M_{200\mathrm{c}} = 10^{13.90} \mathrm{M}_\odot$, which corresponds to the optimal mass at $k = 1.0 \,  {h \,\mathrm{Mpc}^{-1}}$, i.e. $\hat{M}_{k,200\mathrm{c}} (k = 1.0 \, {h \,\mathrm{Mpc}^{-1}}) = 10^{13.90} \mathrm{M}_\odot$. The vertical dashed line indicates the Nyquist frequency of the simulations $k_{\mathrm{Ny}}$. The pink lightly shaded region indicates the scales where our model is not a good indicator of the uncertainty as $k>k_{\mathrm{Ny}}$. \textit{Middle panel:} ratio between the measurements of the suppression in the power spectrum induced by baryons as measured in the hydrodynamical simulations to our best-fitting model for all 400 models. Dashed black lines indicate 1\% accuracy, while dotted lines indicate 2\% accuracy. The solid blue line shows the median ratio, while the blue dashed lines enclose the 15.9 and 84.1 percentile. \textit{Bottom panel:} Mean absolute percentage error for our best-fitting model for all the 400 simulation models.}
\label{fig:model}
\end{figure}

\begin{figure*}
\centering 
\includegraphics[width=0.95\textwidth]{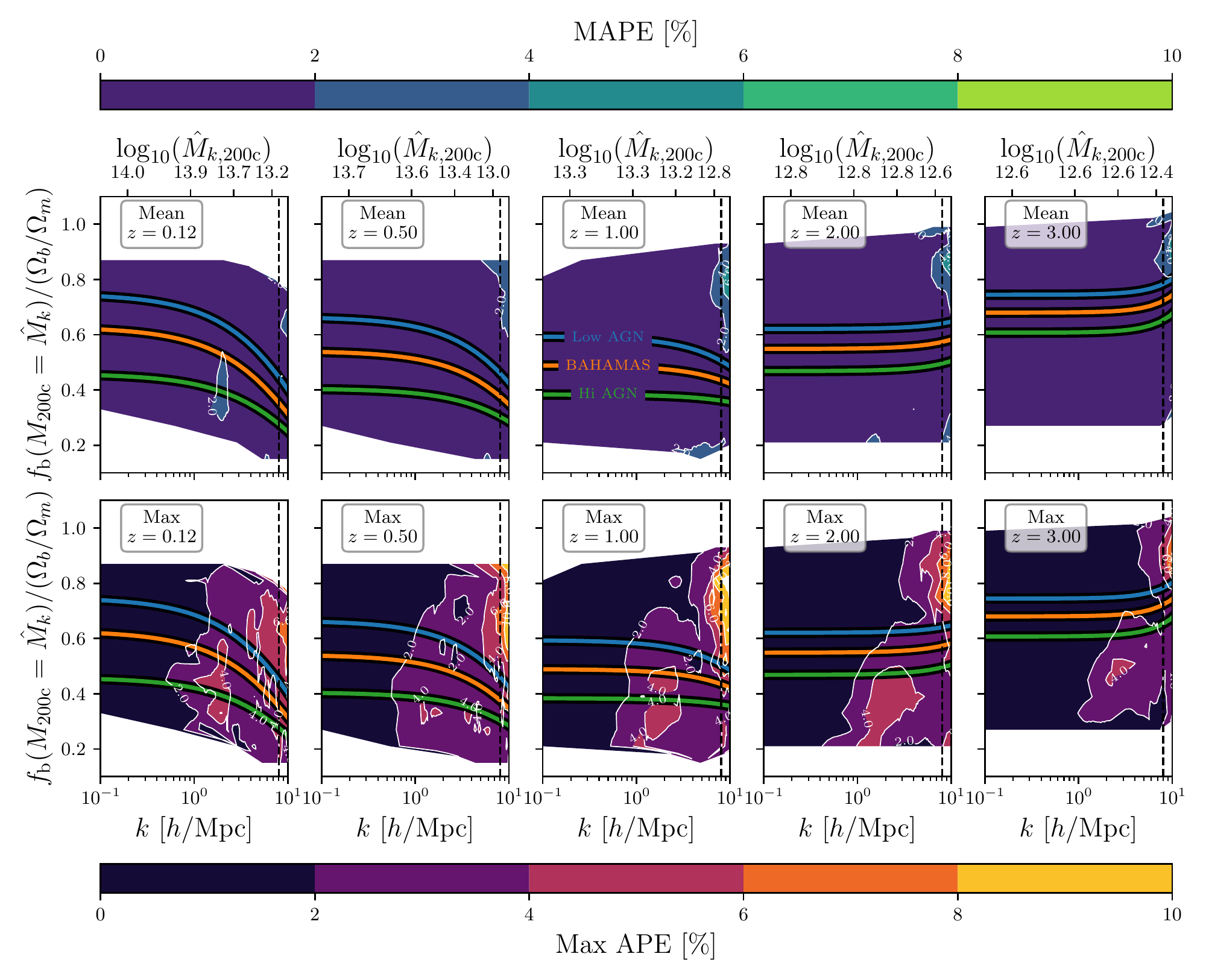}
 \vspace{-1.5em}
 \caption{2D mean \textit{(Top)} and maximum \textit{(Bottom)} absolute percentage error distribution for our best-fitting  model at $z=0.125,0.5, 1.0, 2.0$ and $3.0$. The baryon fraction is computed using the optimal mass of haloes $\hat{M}_{k,200\mathrm{c}}$. The vertical dashed line indicates the Nyquist frequency of the simulations $k_{\mathrm{Ny}}$. $\hat{M}_{k,200\mathrm{c}}$ as function of $k$ and $z$ is calculated using \cref{eq:optimal_mass_fit} and given along the top axis. The three coloured lines indicate the total baryon fraction using the optimal mass for the three BAHAMAS models (Low AGN, fiducial BAHAMAS, and High AGN models. \protect\citealt{mccarthy_bahamas_2017,mccarthy_bahamas_2018}).}
 \label{fig:error}
\end{figure*}

\begin{figure*}
\centering 
\includegraphics[width=0.95\textwidth]{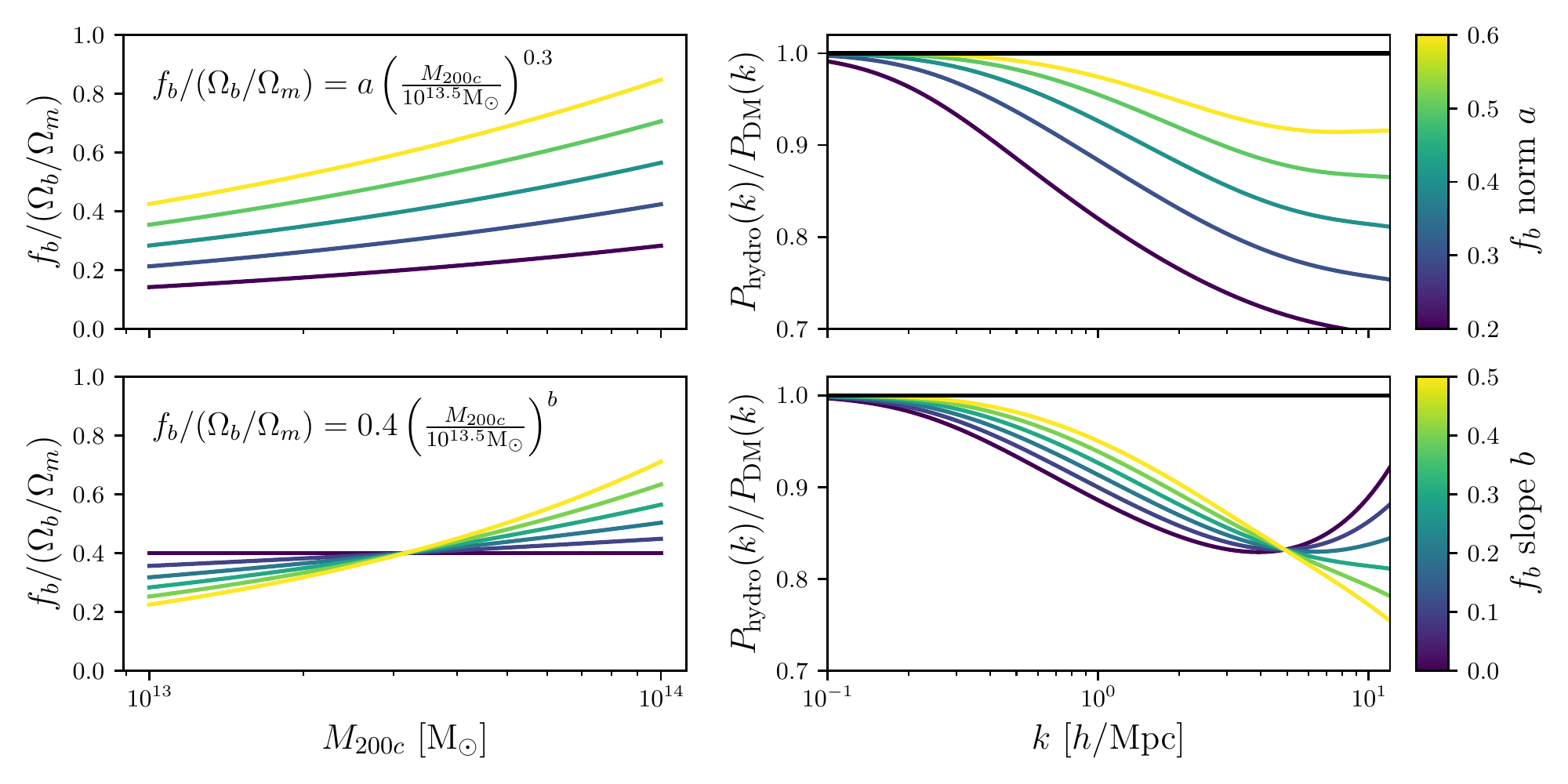}
 \vspace{-1.5em}
 \caption{\textit{Top panel:} Effect of variations in the power-law normalisation ($a$) of the $f_b-M_\mathrm{halo}$ relation (left), as parametrised by \cref{eq:power-law_fb}, on the shape of the suppression of the total matter power spectrum (right) at redshift $z=0.125$. The larger the amount of mass removed from haloes (lower $f_b$), the stronger the suppression on $P(k)$ relative to a baryon complete (DM-only) model. \textit{Bottom panel:} Effect of variations in the power-law slope ($b$).}
 \label{fig:fb_powerlaw}
\end{figure*}

\subsection{Model accuracy}

The level of agreement between the simulations and \texttt{SP(k)} is shown in \cref{fig:model}. To appreciate the performance of the model, in the top panel, we show the fractional impact of baryons on the total matter power spectrum against data for 8 randomly selected models from the 400 simulations at redshift $z=0.1$ spanning a wide range of baryon fractions. Solid lines show the power spectra computed directly from the simulations. Our model is not restrictive to a particular shape of the baryon fraction -- halo mass relation, and in order to show its flexibility, we show in dashed lines our best-fitting model  using binned data for the $f_b-M_\mathrm{halo}$ relation obtained from the simulations, while in dotted lines we use a simple power-law functional form fit to the same relation in the halo mass range of interest (see also \cref{sec:power-law}). 

Whilst the data from the simulations show complex features, the overall behaviour is captured well by our model. By construction, the suppression goes to unity for low values of $k$. The characteristic ``spoon-like'' shape of the power spectrum suppression, the amplitude of which depends on the baryon fraction of haloes, is well captured by the model. In the middle panel we show the ratio between the measurements of the suppression as measured in the hydrodynamical simulations, to our best-fitting model for all 400 models. The solid blue line shows the median ratio as a function of $k$, while the blue dashed lines enclose the 15.9 and 84.1 percentiles (one sigma). In the bottom panel we show the mean absolute percentage error (MAPE), defined as,
\begin{equation}
    \mathrm{MAPE} = \frac{100\%}{n}\sum_{i=1}^n\left|{\frac{y_i - \hat{y_i}}{y_i}}\right|,
\end{equation}
where $y_i$ is the measured variable $P_\mathrm{hydro}(k)/P_\mathrm{DM}(k)$ and $\hat{y}_i$ is the predicted values from our model. The figure shows that the model is able to reproduce the wide range of baryonic effects explored here remarkably well. For all the scale range up to the Nyquist frequency, the MAPE is $\approx 1\%$. For large scales, the model recovers the power spectrum suppression to better than 1\% accuracy, while for the smallest scales considered, the error for most models within $< 5 \%$, with an average of $< 2\%$ for the entire range of baryon fractions. 

We accomplish this significant reduction in the scatter for the power spectrum suppression--baryon fraction relation in our model by selecting the optimal mass that maximises $\rho_s$ for each scale. Nevertheless, for the smallest scales that can be resolved in the simulations (close to the Nyquist frequency, $k_{\mathrm{Ny}} \approx 8 \, {h \,\mathrm{Mpc}^{-1}}$), while the model can recover a mostly monotonic relation, there is still significant scatter around it, as shown with purple dots in \cref{fig:fit}. As the scatter increases with scale $k$, this gives rise to the heteroscedastic behaviour of the errors, i.e. the error in the model increases with scale (dashed blue lines). Nonetheless, our model is an unbiased estimator of the true baryonic effects, as the ratio between the measurements of the suppression as measured in the hydrodynamical simulations to our best-fitting model is centered around unity at all scales.

In \cref{fig:error} we use the individual model errors for the 400 simulations to compute a 2D mean, and maximum, absolute percentage error distribution as a function of scale and baryon fraction for different redshifts. The baryon fraction is computed using the optimal mass of haloes for each corresponding scale and redshift using \cref{eq:optimal_mass_fit}. The mean and maximum errors are computed over linearly spaced baryon fraction bins and logarithmically spaced $k$ bins between $0.1$ and $10.0$ ${h \,\mathrm{Mpc}^{-1}}$. Only pixels with 5 or more data points are shown. The top panels show that our model has a mean error of $\lesssim 2\%$ for the entire scale and redshift range. Furthermore, as shown in the bottom panels of \cref{fig:error}, our model is less accurate for increasing $k$. This can be seen as the increasing contours of maximum absolute percentage error towards the right-hand side of each panel at the bottom of the figure. The maximum absolute percentage error reaches values of $4\%$ to $6\%$ for scales close to the Nyquist frequency at low redshift, and up to $8\%$ to $10\%$ at $z=1-2$. Additionally, the model is less accurate for simulations with extreme feedback prescriptions (low baryon fractions), for scales $k \sim 1 \, {h \,\mathrm{Mpc}^{-1}}$. While a wide range of feedback prescriptions were used to build and calibrate our model, it is important to highlight that for simulations with mean group-scale baryon fractions roughly consistent with observations, the accuracy of our model is at its best, with mean error of ${\approx}1\%$, and a maximum mean error of ${\approx}3\%$ for the whole scale and redshift range. As an example, \cref{fig:error} shows such accuracy around the fiducial BAHAMAS simulation (orange line), which was specifically calibrated to reproduce the galaxy stellar mass function and the observed gas fractions of galaxy groups and clusters at redshift $z \approx 0$ (see \cref{sec:bahamas}),

\subsection{How the baryon fraction shapes the suppression of the total matter power spectrum}\label{sec:power-law}

We now use our model to systematically explore the effect of the $f_b-M_\mathrm{halo}$ relation on the shape of the suppression of the total matter power spectrum. For the mass range that can be relatively well probed in current X-ray and Sunyaev-Zel'dovich effect observations (roughly $10^{13} \lesssim M_{200\mathrm{c}} \,\, [\mathrm{M}_\odot] \lesssim 10^{15} $), the total baryon fraction of haloes can be roughly approximated by a power-law with constant slope \citep[e.g.][]{Mulroy_2019,Akino_2022}. Hence, we use the following parameterisation,
\begin{equation}\label{eq:power-law_fb}
    f_b / (\Omega_b/\Omega_m) = a \left(\frac{M_{200c}}{10^{13.5}\mathrm{M}_\odot} \right)^{b}, 
\end{equation}
where $a$ is the normalisation of the $f_b-M_{200c}$ relation relation at $M_{200c} = {10^{13.5}\mathrm{M}_\odot}$, and $b$ is the power-law slope. 

We use our best fitting parameters for $M_{200c}$ in \cref{tab:opt_mass,tab:sup_fit} and illustrate the effect of changing the power-law fitting parameters $a$ and $b$ at redshift $z=0.125$ in \cref{fig:fb_powerlaw}. The top panel shows variations in the power-law normalisation while keeping $b=0.3$. The bottom panel shows variations in the power-law slope while keeping $a=0.4$. The parameters $a=0.4$ and $b=0.3$ were selected as they are roughly consistent with the baryon fraction - halo mass relation for the fiducial BAHAMAS simulation at the corresponding redshift. 

Consistent with \citet{van_daalen_2020}, the top panel shows the large effect of the normalisation of the baryon fraction - halo mass relation on the suppression of the matter power spectrum. As mentioned above, the effect of baryon physics on the matter power spectrum is fundamentally a gravitational process. Hence, the larger the amount of mass removed from haloes (lower $f_b$), the stronger the suppression on $P(k)$ relative to a baryon complete (DM-only) model. Furthermore, very low baryon fractions (associated with stronger AGN feedback) can have a non-negligible impact the power spectrum suppression on very large scales ($ k {\sim} 0.1\,\, h\,\mathrm{Mpc}^{-1}$, dark-blue line). 

In the bottom panel, we show that the slope of the baryon fraction - halo mass relation has a large effect on the scale of the minimum of the ``spoon-like'' shape of the power spectrum suppression. 

\subsection{Alternative approaches to fitting the suppression}

Our approach above was to apply simple parametric forms to describe the correlation between the power spectrum suppression and the baryon fraction of haloes that contribute most significantly at a given scale.  We have shown that, in spite of its simplicity, the model provides a very accurate description of our large simulation data set.  Nevertheless, our approach to modelling the simulation data is not unique and there may be alternative (e.g., non-parametric) approaches that could lead to even more accurate results.  For example, principle component analysis (PCA) applied to simulation power spectra has been used as a way of characterising the impact of baryons and allowing for their marginalisation in cosmological pipelines (e.g., \citealt{eifler_2015}).  A related strategy is to model the principle components via Gaussian Processes (e.g., \citealt{heitmann_2014}) or sparse polynomial chaos expansion (e.g., \citealt{euclid_2019}), although to our knowledge these methods have so far been applied to only the non-linear correction to the matter power spectrum and not on the impact of baryons. In addition, neural networks have recently been used to construct a baryon emulator by \citet{arico_2021}. The neural net was trained on a large sample of power spectra generated by the baryonification approach.

While there are various advantages and disadvantages to the alternative approaches described above, we have elected to use the approach described above because it is more intuitive in terms of visualising the correlations and it allows us to more easily enforce physically-motivated boundary conditions, such as requiring that suppression of the power spectrum goes to unity as ${k {\to} 0}$ (very large scales). That being said, we have also experimented with both a Gaussian Process-based emulator and a simple neural network using the baryon fractions as the input quantities. We find that these non-parametric approaches do not significantly improve on the accuracy of our fiducial analytic method and we therefore elected to retain the latter as our main model.

\begin{figure*}
\centering 
\includegraphics[width=0.95\textwidth]{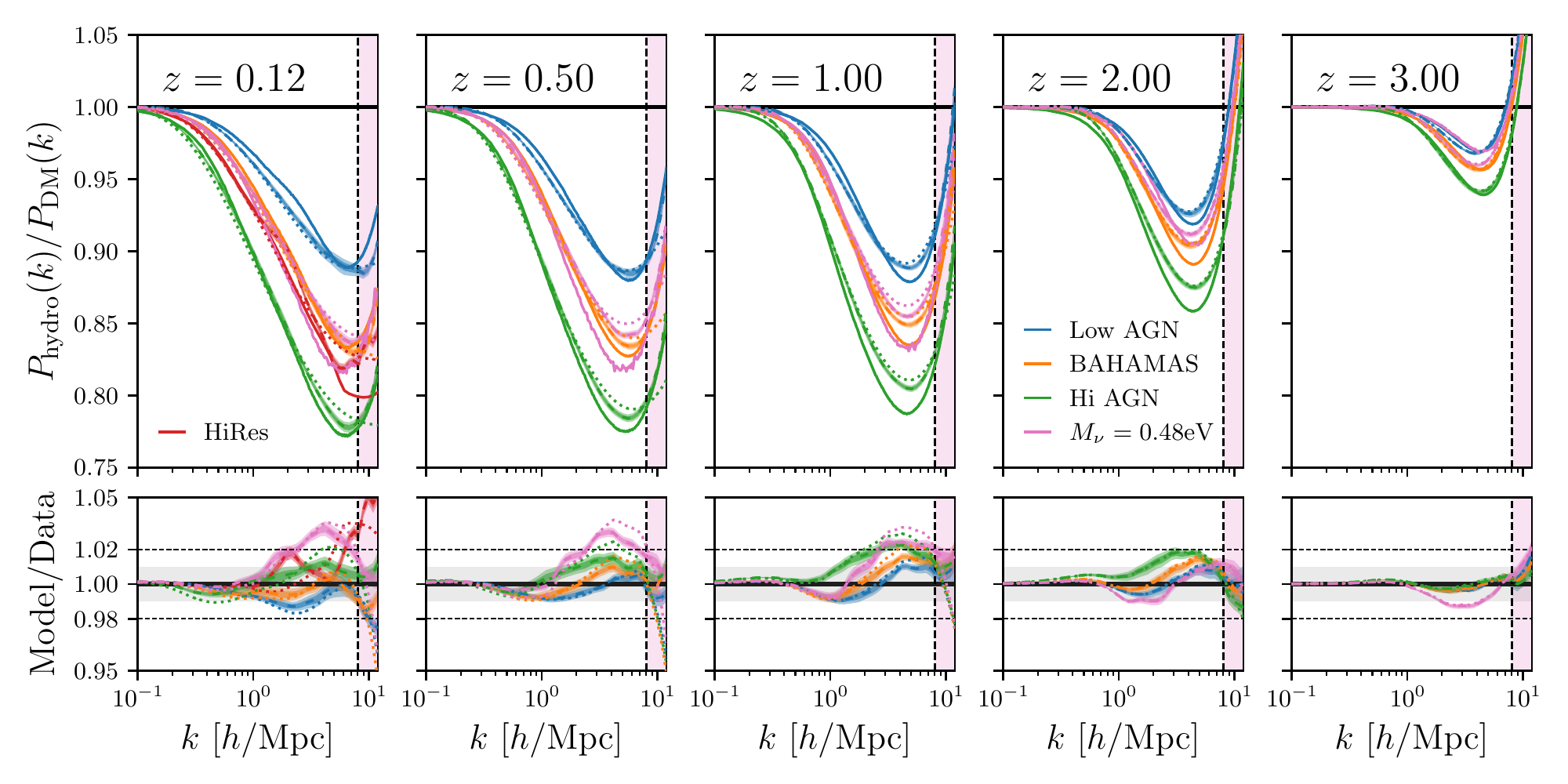}
 \vspace{-1.5em}
 \caption{Fractional impact of baryons on the total matter power spectrum for three BAHAMAS models (Low AGN, fiducial, and High AGN \protect\citealt{mccarthy_bahamas_2017,mccarthy_bahamas_2018}), the high resolution (HiRes  \protect\citealt{mccarthy_bahamas_2017}) and massive neutrino ($M_\nu = 0.48$ eV \protect\citealt{mummery_separate_2017,mccarthy_bahamas_2018}) variations. Solid lines are the power spectra computed the hydrodynamical simulations. Dashed lines represent the best-fitting model obtained with \texttt{SP(k)} using binned data for the $f_b-M_\mathrm{halo}$ relation. Dotted lines use a simple power-law fit to the total baryon fraction. The light shaded regions enclose the 68\% and 95\% confidence interval from statistical errors (only shown for binned data). The vertical dashed line indicates the Nyquist frequency of the ANTILLES simulations $k_{\mathrm{Ny}}$. The pink lightly shaded region indicates the scales where our model is not a good indicator of the uncertainty as $k>k_{\mathrm{Ny}}$. \textit{Bottom panel:} ratio between the measurements of the suppression in the power spectrum induced by baryons as measured in the hydrodynamical simulations to our best-fitting model. The grey band highlights 1\% accuracy, while dashed lines indicate 2\% accuracy.}
 \label{fig:bahamas}
\end{figure*}

\section{Testing against BAHAMAS}\label{sec:bahamas}

We employ the BAHAMAS suite of cosmological hydrodynamical simulations \citep{mccarthy_bahamas_2017,mccarthy_bahamas_2018} to test the accuracy of our model using simulations outside our calibration set. The reference BAHAMAS simulation uses the same cosmology and particle mass resolution as the simulations used in this study, but in a much larger box. The fiducial BAHAMAS run consist of a comoving volume with ${400 \,\, \mathrm{cMpc} \, h^{-1}}$ on a side and $2 \times 1024^3$ particles of $m_\mathrm{DM} = 3.85 \times 10^9 \mathrm{M}_\odot \, h^{-1}$ dark matter particle mass, and $m_\mathrm{g} = 7.66 \times 10^8 \mathrm{M}_\odot \, h^{-1}$ initial gas particle mass. A full discussion of the sub-grid implementation, including the prescriptions for star formation, gas heating and cooling, BH formation and supernovae and AGN feedback models can be found in \citet{mccarthy_bahamas_2017} (see also \citealt{schaye_physics_2010}). As the BAHAMAS suite of simulations aim to study the impact of baryonic process on LSS, it was of paramount importance to ensure that group size and more massive haloes had the correct baryon fractions, as these haloes contribute the most to the matter power spectrum. Therefore, the BAHAMAS approach was to calibrate the efficiencies of the stellar and AGN feedback to reproduce galaxy stellar mass function and the observed gas fractions of galaxy groups and clusters at redshift $z{\approx}0$. We utilise three different variations of the BAHAMAS simulations that differ only in the strength of their AGN feedback prescription. This strength is determined by the AGN sub-grid heating temperature, which is the temperature increase given to gas particles associated to each AGN feedback event. 

Additionally, as a ``proof of concept'', we test our model against two additional simulations, one that varies the mass resolution, but is otherwise identical to the fiducial BAHAMAS model, and one that changes the cosmology by including the effects of massive neutrinos, both on the background expansion rate and the growth of density fluctuations (see further discussion in \cref{sec:discussion}). For the high resolution study, we used the ``high res'' model presented in \cite{mccarthy_bahamas_2017}, which left the subgrid feedback parameters unchanged with respect to the BAHAMAS fiducial resolution model, but it has 8 times better mass resolution (comoving volume of ${100 \,\, \mathrm{cMpc} \, h^{-1}}$ on a side and $2 \times 512^3$ particles). This represents a ‘strong’ convergence test in the terminology of \cite{schaye_eagle_2015}. We note that only the particle data at redshifts $z=0$ and $z=0.125$ for these runs is still available. Hence, we limited our comparison for the ``high res'' run to redshift $z=0.125$.

We have also used our model to predict the baryonic suppression for the most extreme massive neutrino variation of the WMAP9 cosmology in the BAHAMAS suite \citep{mummery_separate_2017,mccarthy_bahamas_2018}. This model left the subgrid feedback parameters unchanged with respect to the fiducial BAHAMAS model, but uses a total summed neutrino mass of $M_\nu = 0.48$ eV. For this model, the amplitude of the density fluctuations at the epoch of recombination $A_s$, as inferred by WMAP9 data assuming massless neutrinos, was held fixed in order to retain agreement with the observed CMB angular power spectrum, and the value of $\sigma_8$ was adjusted accordingly. Additionally, the matter density parameter of cold dark matter was adjusted to maintain a flat universe model. All other cosmological parameters are held fixed.

In \cref{fig:bahamas} we show the impact of baryons on the total matter power spectrum for the three BAHAMAS models (Low AGN, fiducial, and High AGN), the high resolution (HiRes) and massive neutrino ($M_\nu = 0.48$ eV) variations. In order to compute the baryon fraction from the simulations to use in \cref{eq:opt_fb}, we used a piecewise cubic Hermite interpolating polynomial using narrow bins of $M_{200\mathrm{c}}$ with 0.1 dex width. This approach provides a high degree of flexibility, as simulations show evidence for a slightly mass-dependent slope \citep{Farahi_2018}. Additionally, we show in dotted lines the results using power law fits to the median baryon fraction in the halo mass range of interest at each redshift. 

The figure shows that our model follows the simulation response typically to within $1\%$, with a maximum error of a $\sim 3\%$ across the entire range of scales and redshifts shown. The errors at all scales and redshifts are consistent with the expected errors from the mean error contours in \cref{fig:error}. The largest errors are at $2 \lesssim k \,\, [h\,\mathrm{Mpc}^{-1}] \lesssim 5$ at $z=1$ for most models. In particular, for the lowest baryon fraction model, as expected from the maximum error contours in \cref{fig:error}. For the massive neutrino cosmology, the largest errors are at $2 \lesssim k \,\, [h\,\mathrm{Mpc}^{-1}] \lesssim 5$ at $z=0.5$ and $z=0.1$. By construction, the model recovers the suppression (or rather, lack thereof) at the large scales. Hence, the errors are very small at the largest of scales. Finally, we do not find a significant decrease of the performance of the model when using a simple power law, compared to a more precise interpolation of the baryon fraction as a function of halo mass. 

\section{Discussion}\label{sec:discussion}

It is important to note that our model was built from a suite of simulations that widely varied the feedback parameters but only within a single cosmology (i.e., the cosmological parameters were fixed). How worried should we be about possible degeneracies between astrophysics in cosmology? In other words, does the scaling derived here apply to other cosmologies?  Our previous work based on the BAHAMAS simulations has shown that the effects of baryon physics on the matter distribution appear to be separable from (independent of) variations in the cosmological parameters, including extensions to massive neutrinos \citep{mummery_separate_2017}, dynamical dark energy \citep{pfeifer_2020}, and a running of scalar spectral index \citep{stafford_2020}. In those studies, we demonstrated that the resulting cluster baryon fractions and the suppression of the matter power spectrum match those of our fiducial cosmology model to typically percent level accuracy or better.  This is consistent with the findings of previous work based on the baryonification formalism, which has also shown that the suppression of the power spectrum due to baryons is independent of variations in the parameters of the standard cosmological model to approximately the percent level, with the exception of a clear dependence on the universal baryon fraction, $\Omega_b/\Omega_m$ \citep{schneider_2020,arico_2021}.

The dependence on the universal baryon fraction is straightforward to interpret: in the absence of feedback, groups and clusters are expected to contain a baryon fraction that reflects that of the universe as a whole \citep{white_1993}, thus a higher (lower) universal fraction requires more (less) efficient feedback to obtain a particular (e.g., observed) baryon fraction in groups/clusters.  In other words, if the baryon fractions of groups/clusters are used as an independent constraint on the feedback-induced suppression of the matter power spectrum, one must also specify the universal baryon fraction that is assumed.  This is why our model takes as its independent variable the halo baryon fraction normalised by the universal baryon fraction.  In practice, the quantity $\Omega_b/\Omega_m$ is precisely pinned down by observations of the CMB, such that the uncertainties in baryon fractions of haloes are dominated by the statistical and systematic errors in the halo (total and baryon) mass measurements.  Nevertheless, marginalising over the uncertainties in both the halo and universal baryon fractions is recommended.  Going forward, it will be important to continue to check for potential degeneracies between baryon and cosmological physics as we extend beyond the standard model of cosmology and also explore other cosmological observables (e.g., galaxy clustering, cluster counts, and so on).

Another caveat of our study is that, while we have explored a wide range of the feedback parameter volume in the present study, this has only been done within the context of a given framework (or parametrisation) for feedback.  It will be important for other groups to independently test and perhaps refine the predictions for the impact of baryon physics on the matter power spectrum as new simulations become available. The study of \citet{van_daalen_2020}, which compared a limited number of simulations in the literature but which spanned a wide range of resolutions, hydro solvers, and feedback frameworks, is very suggestive that the physics of $P(k)$ is mainly driven by the baryon fractions.  However, these simulations may not explore the full range of possibilities (see e.g., \citealt{debackere_impact_2020,amon_2022}), so it is important to continue exploring the effects of baryons using a wide variety of methods, including the halo model, the baryonification formalism, and different hydrodynamical simulations. 

As an initial ``proof of concept'', we test the separability of feedback parameters (and resolution effects) from the cosmological model in \cref{fig:bahamas} using a higher resolution simulation and one that changes the cosmology by including the effects of massive neutrinos. While this is a limited test, we find that our model based solely on the baryon fraction recovers the suppression of the matter power spectrum to a typically better than $\sim 2 \%$ accuracy. Further investigation including different feedback prescriptions, cosmological models, and resolution effects will be thoroughly explored in a follow-up paper. 

\section{Summary and conclusions}\label{sec:conclusion}

We have presented ANTILLES, a new suite of 400 hydrodynamical simulations that explores the ``feedback landscape'' associated with baryon physics.  We have shown that a relatively simple yet remarkably accurate model can be constructed for the suppression of the matter power spectrum using only the mean baryon fraction of haloes (specifically the baryon fraction for haloes that dominate the power spectrum at a given scale, or wavenumber).  Our work follows on from the recent study of \citet{van_daalen_2020}, by expanding greatly the number of simulations used to map this relationship and by pushing to a wider range of scales and redshifts, which are requirements for current and upcoming large-scale structure surveys including cosmic shear.

The main specific findings of our study may be summarised as follows:
\begin{itemize}
    \item In agreement with \citet{van_daalen_2020}, we find that the fractional impact of baryon physics on the present-day non-linear matter power spectrum up to $k \sim 1$ $h$ Mpc$^{-1}$ correlates very strongly with the baryon fraction of haloes with total masses of $\sim10^{14}$ M$_\odot$ (see \Cref{fig:PS}). At smaller scales, however, the relation weakens somewhat, as also found by \citet{van_daalen_2020}.
    \item The weakening in the relation between the power spectrum suppression and the baryon fraction at a mass scale of $\sim10^{14}$ M$_\odot$ is a result of lower mass haloes contributing more significantly at smaller scales (see also \citealt{van_daalen_contributions_2015,mead_hydrodynamical_2020}). Using the simulations, we have empirically determined the halo mass scale whose baryon fraction correlates most strongly with the suppression of the power spectrum as a function of both scale and redshift (see \cref{fig:spearman} and \cref{fig:best_mass}).  We refer to this mass scale as the `optimal' mass.
    \item Our $P(k)$ suppression model, called \texttt{SP(k)}, is constructed by fitting a simple parametric form to the power spectrum suppression as a function of scale ($k$), redshift ($z$), and the normalised baryon fraction at the optimal scale. See \cref{eq:sup_fit}.
    \item We characterise the error in our best-fit model relative to our simulation training set in \cref{fig:model} and \cref{fig:error}, showing our model to be accurate to typically better than ${\approx}2\%$ percent accuracy over the range $0.1 \lesssim k \ [h \ \textrm{Mpc}^{-1}] \lesssim 10$ and from $z =0.1$--3. 
    \item We tested our model against an independent set of BAHAMAS simulations \citep{mccarthy_bahamas_2017,mccarthy_bahamas_2018}, including a high resolution box (\citealt{mccarthy_bahamas_2017}) and massive neutrino cosmology ($M_\nu = 0.48$ eV, \citealt{mummery_separate_2017,mccarthy_bahamas_2018}), showing a similar level of accuracy (see \cref{fig:bahamas}).
    \item We make a Python implementation of our model publicly available at \href{https://github.com/jemme07/pyspk}{https://github.com/jemme07/pyspk}
\end{itemize}

With a fast, accurate model for characterising the effects of baryon physics on the non-linear matter power spectrum, it is possible to straightforwardly incorporate these effects in existing theoretical pipelines based on gravity-only calculations (e.g., emulators for the absolute non-linear power spectrum or the halo model). This will allow one to consistently propagate the uncertainties in astrophysics (due to our uncertainties in the baryon fractions of groups/clusters as a function of mass and redshift) through to the cosmological constraints. An advantage that our model has over existing methods based on the halo model or the `baryonification' formalism is that it effectively depends only on a single physically-meaningful parameter (the baryon fraction). The benefit of this is that observational constraints on the baryon fraction could be used to inform the priors used in cosmological analysis, which in turn should help minimise the degradation of cosmological constraints that results from marginalising over baryon effects \citep[e.g.][]{amon_2022}. Furthermore, by providing a set of analytic equations, the model can be easily ``inverted'' and allows for rapid experiments to be conducted, providing a powerful tool to explore the differential effects of baryonic physics. 

Finally, as our model is expressed in terms of the baryon fraction of groups and clusters, we require observational constraints to either inform the priors that will be used in cosmological pipelines or to be used directly in a joint analysis. While the baryon fractions are pinned down reasonably precisely in the low-redshift universe (perhaps $z \lesssim 0.2$), there are currently very few constraints at higher redshifts for haloes with $M \lesssim 10^{14}$ M$_\odot$. The situation may soon change, though, as eROSITA (X-ray), Advanced ACT (tSZ, kSZ) and Simons Observatory (tSZ, kSZ) data of large numbers of high-$z$ groups starts to become available.  An important consideration in these analyses will be carefully modelling the selection function to enable an unbiased measurement of the group baryon fractions. Analyses of mock X-ray and SZ observations of cosmological hydrodynamical simulations will be important in this endeavour.  

\section*{Acknowledgements}
The authors would like to thank Dr. Alexandra Amon for her thorough report and valuable comments as the referee, which greatly contributed to the improvement of the paper. This project has received funding from the European Research Council (ERC) under the European Union’s Horizon 2020 research and innovation programme (grant agreement No 769130). This work was supported by the Science and Technology Facilities Council [ST/P000541/1]. This work used the DiRAC@Durham facility managed by the Institute for Computational Cosmology on behalf of the STFC DiRAC HPC Facility (www.dirac.ac.uk). The equipment was funded by BEIS capital funding via STFC capital grants ST/K00042X/1, ST/P002293/1 and ST/R002371/1, Durham University and STFC operations grant ST/R000832/1. DiRAC is part of the National e-Infrastructure.

\section*{Data availability}
The data underlying this article may be shared on reasonable request to the corresponding author.

%%%%%%%%%%%%%%%%%%%%%%%%%%%%%%%%%%%%%%%%%%%%%%%%%%



\bibliographystyle{mnras}
\bibliography{newbib} % if your bibtex file is called example.bib

\begin{thebibliography}{}
\makeatletter
\relax
\def\mn@urlcharsother{\let\do\@makeother \do\$\do\&\do\#\do\^\do\_\do\%\do\~}
\def\mn@doi{\begingroup\mn@urlcharsother \@ifnextchar [ {\mn@doi@}
  {\mn@doi@[]}}
\def\mn@doi@[#1]#2{\def\@tempa{#1}\ifx\@tempa\@empty \href
  {http://dx.doi.org/#2} {doi:#2}\else \href {http://dx.doi.org/#2} {#1}\fi
  \endgroup}
\def\mn@eprint#1#2{\mn@eprint@#1:#2::\@nil}
\def\mn@eprint@arXiv#1{\href {http://arxiv.org/abs/#1} {{\tt arXiv:#1}}}
\def\mn@eprint@dblp#1{\href {http://dblp.uni-trier.de/rec/bibtex/#1.xml}
  {dblp:#1}}
\def\mn@eprint@#1:#2:#3:#4\@nil{\def\@tempa {#1}\def\@tempb {#2}\def\@tempc
  {#3}\ifx \@tempc \@empty \let \@tempc \@tempb \let \@tempb \@tempa \fi \ifx
  \@tempb \@empty \def\@tempb {arXiv}\fi \@ifundefined
  {mn@eprint@\@tempb}{\@tempb:\@tempc}{\expandafter \expandafter \csname
  mn@eprint@\@tempb\endcsname \expandafter{\@tempc}}}

\bibitem[\protect\citeauthoryear{{Acuto}, {McCarthy}, {Kwan}, {Salcido},
  {Stafford}  \& {Font}}{{Acuto} et~al.}{2021}]{acuto_2021}
{Acuto} A.,  {McCarthy} I.~G.,  {Kwan} J.,  {Salcido} J.,  {Stafford} S.~G.,
  {Font} A.~S.,  2021, \mn@doi [Monthly Notices of the Royal Astronomical
  Society] {10.1093/mnras/stab2834}, \href
  {https://ui.adsabs.harvard.edu/abs/2021MNRAS.508.3519A} {508, 3519}

\bibitem[\protect\citeauthoryear{{Akino} et~al.,}{{Akino}
  et~al.}{2022}]{Akino_2022}
{Akino} D.,  et~al., 2022, \mn@doi [Publications of the Astronomical Society of
  Japan] {10.1093/pasj/psab115}, \href
  {https://ui.adsabs.harvard.edu/abs/2022PASJ..tmp....2A} {}

\bibitem[\protect\citeauthoryear{{Amon} \& {Efstathiou}}{{Amon} \&
  {Efstathiou}}{2022}]{amon_2022}
{Amon} A.,  {Efstathiou} G.,  2022, \mn@doi [Monthly Notices of the Royal
  Astronomical Society] {10.1093/mnras/stac2429}, \href
  {https://ui.adsabs.harvard.edu/abs/2022MNRAS.516.5355A} {516, 5355}

\bibitem[\protect\citeauthoryear{{Amon} et~al.,}{{Amon}
  et~al.}{2022}]{Amon2022}
{Amon} A.,  et~al., 2022, \mn@doi [\prd] {10.1103/PhysRevD.105.023514}, \href
  {https://ui.adsabs.harvard.edu/abs/2022PhRvD.105b3514A} {105, 023514}

\bibitem[\protect\citeauthoryear{{Angulo}, {Zennaro}, {Contreras}, {Aric{\`o}},
  {Pellejero-Iba{\~n}ez}  \& {St{\"u}cker}}{{Angulo}
  et~al.}{2021}]{angulo_2021}
{Angulo} R.~E.,  {Zennaro} M.,  {Contreras} S.,  {Aric{\`o}} G.,
  {Pellejero-Iba{\~n}ez} M.,   {St{\"u}cker} J.,  2021, \mn@doi [Monthly
  Notices of the Royal Astronomical Society] {10.1093/mnras/stab2018}, \href
  {https://ui.adsabs.harvard.edu/abs/2021MNRAS.507.5869A} {507, 5869}

\bibitem[\protect\citeauthoryear{{Aric{\`o}}, {Angulo}, {Contreras},
  {Ondaro-Mallea}, {Pellejero-Iba{\~n}ez}  \& {Zennaro}}{{Aric{\`o}}
  et~al.}{2021}]{arico_2021}
{Aric{\`o}} G.,  {Angulo} R.~E.,  {Contreras} S.,  {Ondaro-Mallea} L.,
  {Pellejero-Iba{\~n}ez} M.,   {Zennaro} M.,  2021, \mn@doi [Monthly Notices of
  the Royal Astronomical Society] {10.1093/mnras/stab1911}, \href
  {https://ui.adsabs.harvard.edu/abs/2021MNRAS.506.4070A} {506, 4070}

\bibitem[\protect\citeauthoryear{{Aric{\`o}}, {Angulo}, {Zennaro}, {Contreras},
  {Chen}  \& {Hern{\'a}ndez-Monteagudo}}{{Aric{\`o}} et~al.}{2023}]{arico_2023}
{Aric{\`o}} G.,  {Angulo} R.~E.,  {Zennaro} M.,  {Contreras} S.,  {Chen} A.,
  {Hern{\'a}ndez-Monteagudo} C.,  2023, \mn@doi [arXiv e-prints]
  {10.48550/arXiv.2303.05537}, \href
  {https://ui.adsabs.harvard.edu/abs/2023arXiv230305537A} {p. arXiv:2303.05537}

\bibitem[\protect\citeauthoryear{{Asgari} et~al.,}{{Asgari}
  et~al.}{2021}]{Asgari2021}
{Asgari} M.,  et~al., 2021, \mn@doi [\aap] {10.1051/0004-6361/202039070}, 645,
  A104

\bibitem[\protect\citeauthoryear{Baldry et~al.,}{Baldry
  et~al.}{2012}]{baldry_galaxy_2012}
Baldry I.~K.,  et~al., 2012, \mn@doi [Monthly Notices of the Royal Astronomical
  Society] {10.1111/j.1365-2966.2012.20340.x}, 421, 621

\bibitem[\protect\citeauthoryear{{Balogh}, {McCarthy}, {Bower}  \&
  {Eke}}{{Balogh} et~al.}{2008}]{balogh_2008}
{Balogh} M.~L.,  {McCarthy} I.~G.,  {Bower} R.~G.,   {Eke} V.~R.,  2008,
  \mn@doi [Monthly Notices of the Royal Astronomical Society]
  {10.1111/j.1365-2966.2008.12903.x}, \href
  {https://ui.adsabs.harvard.edu/abs/2008MNRAS.385.1003B} {385, 1003}

\bibitem[\protect\citeauthoryear{Bernardi, Meert, Sheth, Vikram,
  Huertas-Company, Mei  \& Shankar}{Bernardi
  et~al.}{2013}]{bernardi_massive_2013}
Bernardi M.,  Meert A.,  Sheth R.~K.,  Vikram V.,  Huertas-Company M.,  Mei S.,
    Shankar F.,  2013, \mn@doi [Monthly Notices of the Royal Astronomical
  Society] {10.1093/mnras/stt1607}, 436, 697

\bibitem[\protect\citeauthoryear{{Bond}, {Efstathiou}  \& {Silk}}{{Bond}
  et~al.}{1980}]{Bond_80}
{Bond} J.~R.,  {Efstathiou} G.,   {Silk} J.,  1980, \mn@doi [\prl]
  {10.1103/PhysRevLett.45.1980}, \href
  {https://ui.adsabs.harvard.edu/abs/1980PhRvL..45.1980B} {45, 1980}

\bibitem[\protect\citeauthoryear{Booth \& Schaye}{Booth \&
  Schaye}{2009}]{booth_cosmological_2009}
Booth C.~M.,  Schaye J.,  2009, \mn@doi [Monthly Notices of the Royal
  Astronomical Society] {10.1111/j.1365-2966.2009.15043.x}, 398, 53

\bibitem[\protect\citeauthoryear{{Castro}, {Borgani}, {Dolag}, {Marra},
  {Quartin}, {Saro}  \& {Sefusatti}}{{Castro} et~al.}{2021}]{castro_2021}
{Castro} T.,  {Borgani} S.,  {Dolag} K.,  {Marra} V.,  {Quartin} M.,  {Saro}
  A.,   {Sefusatti} E.,  2021, \mn@doi [Monthly Notices of the Royal
  Astronomical Society] {10.1093/mnras/staa3473}, \href
  {https://ui.adsabs.harvard.edu/abs/2021MNRAS.500.2316C} {500, 2316}

\bibitem[\protect\citeauthoryear{{Chen} et~al.,}{{Chen}
  et~al.}{2023}]{chen_2023}
{Chen} A.,  et~al., 2023, \mn@doi [\mnras] {10.1093/mnras/stac3213}, \href
  {https://ui.adsabs.harvard.edu/abs/2023MNRAS.518.5340C} {518, 5340}

\bibitem[\protect\citeauthoryear{Chisari et~al.,}{Chisari
  et~al.}{2018}]{chisari_impact_2018}
Chisari N.~E.,  et~al., 2018, \mn@doi [Monthly Notices of the Royal
  Astronomical Society] {10.1093/mnras/sty2093}, 480, 3962

\bibitem[\protect\citeauthoryear{{Chisari} et~al.,}{{Chisari}
  et~al.}{2019}]{chisari_modelling_2019}
{Chisari} N.~E.,  et~al., 2019, \mn@doi [The Open Journal of Astrophysics]
  {10.21105/astro.1905.06082}, \href
  {https://ui.adsabs.harvard.edu/abs/2019OJAp....2E...4C} {2, 4}

\bibitem[\protect\citeauthoryear{Cooray \& Sheth}{Cooray \&
  Sheth}{2002}]{cooray_halo_2002}
Cooray A.,  Sheth R.,  2002, \mn@doi [\physrep]
  {10.1016/S0370-1573(02)00276-4}, 372, 1

\bibitem[\protect\citeauthoryear{Cullen \& Dehnen}{Cullen \&
  Dehnen}{2010}]{cullen_inviscid_2010}
Cullen L.,  Dehnen W.,  2010, \mn@doi [Monthly Notices of the Royal
  Astronomical Society] {10.1111/j.1365-2966.2010.17158.x}, 408, 669

\bibitem[\protect\citeauthoryear{Dalla~Vecchia \& Schaye}{Dalla~Vecchia \&
  Schaye}{2008}]{dalla_vecchia_simulating_2008}
Dalla~Vecchia C.,  Schaye J.,  2008, \mn@doi [Monthly Notices of the Royal
  Astronomical Society] {10.1111/j.1365-2966.2008.13322.x}, 387, 1431

\bibitem[\protect\citeauthoryear{Davis, Efstathiou, Frenk  \& White}{Davis
  et~al.}{1985}]{davis_evolution_1985}
Davis M.,  Efstathiou G.,  Frenk C.~S.,   White S. D.~M.,  1985, \mn@doi [The
  Astrophysical Journal] {10.1086/163168}, 292, 371

\bibitem[\protect\citeauthoryear{{DeRose} et~al.,}{{DeRose}
  et~al.}{2019}]{derose_2019}
{DeRose} J.,  et~al., 2019, \mn@doi [The Astrophysical Journal]
  {10.3847/1538-4357/ab1085}, \href
  {https://ui.adsabs.harvard.edu/abs/2019ApJ...875...69D} {875, 69}

\bibitem[\protect\citeauthoryear{Debackere, Schaye  \& Hoekstra}{Debackere
  et~al.}{2020}]{debackere_impact_2020}
Debackere S. N.~B.,  Schaye J.,   Hoekstra H.,  2020, \mn@doi [Monthly Notices
  of the Royal Astronomical Society] {10.1093/mnras/stz3446}, 492, 2285

\bibitem[\protect\citeauthoryear{{Debackere}, {Schaye}  \&
  {Hoekstra}}{{Debackere} et~al.}{2021}]{debackere_2021}
{Debackere} S. N.~B.,  {Schaye} J.,   {Hoekstra} H.,  2021, \mn@doi [Monthly
  Notices of the Royal Astronomical Society] {10.1093/mnras/stab1326}, \href
  {https://ui.adsabs.harvard.edu/abs/2021MNRAS.505..593D} {505, 593}

\bibitem[\protect\citeauthoryear{Deutsch \& Deutsch}{Deutsch \&
  Deutsch}{2012}]{DEUTSCH2012763}
Deutsch J.~L.,  Deutsch C.~V.,  2012, \mn@doi [Journal of Statistical Planning
  and Inference] {https://doi.org/10.1016/j.jspi.2011.09.016}, 142, 763

\bibitem[\protect\citeauthoryear{{Driver} et~al.,}{{Driver}
  et~al.}{2022}]{2022MNRAS.513..439D}
{Driver} S.~P.,  et~al., 2022, \mn@doi [Monthly Notices of the Royal
  Astronomical Society] {10.1093/mnras/stac472}, \href
  {https://ui.adsabs.harvard.edu/abs/2022MNRAS.513..439D} {513, 439}

\bibitem[\protect\citeauthoryear{Durier \& Dalla~Vecchia}{Durier \&
  Dalla~Vecchia}{2012}]{durier_implementation_2012}
Durier F.,  Dalla~Vecchia C.,  2012, \mn@doi [Monthly Notices of the Royal
  Astronomical Society] {10.1111/j.1365-2966.2011.19712.x}, 419, 465

\bibitem[\protect\citeauthoryear{{Eifler}, {Krause}, {Dodelson}, {Zentner},
  {Hearin}  \& {Gnedin}}{{Eifler} et~al.}{2015}]{eifler_2015}
{Eifler} T.,  {Krause} E.,  {Dodelson} S.,  {Zentner} A.~R.,  {Hearin} A.~P.,
  {Gnedin} N.~Y.,  2015, \mn@doi [Monthly Notices of the Royal Astronomical
  Society] {10.1093/mnras/stv2000}, \href
  {https://ui.adsabs.harvard.edu/abs/2015MNRAS.454.2451E} {454, 2451}

\bibitem[\protect\citeauthoryear{{Euclid Collaboration} et~al.,}{{Euclid
  Collaboration} et~al.}{2019}]{euclid_2019}
{Euclid Collaboration} et~al., 2019, \mn@doi [Monthly Notices of the Royal
  Astronomical Society] {10.1093/mnras/stz197}, \href
  {https://ui.adsabs.harvard.edu/abs/2019MNRAS.484.5509E} {484, 5509}

\bibitem[\protect\citeauthoryear{{Farahi}, {Evrard}, {McCarthy}, {Barnes}  \&
  {Kay}}{{Farahi} et~al.}{2018}]{Farahi_2018}
{Farahi} A.,  {Evrard} A.~E.,  {McCarthy} I.,  {Barnes} D.~J.,   {Kay} S.~T.,
  2018, \mn@doi [Monthly Notices of the Royal Astronomical Society]
  {10.1093/mnras/sty1179}, \href
  {https://ui.adsabs.harvard.edu/abs/2018MNRAS.478.2618F} {478, 2618}

\bibitem[\protect\citeauthoryear{{Gonzalez}, {Sivanandam}, {Zabludoff}  \&
  {Zaritsky}}{{Gonzalez} et~al.}{2013}]{2013ApJ...778...14G}
{Gonzalez} A.~H.,  {Sivanandam} S.,  {Zabludoff} A.~I.,   {Zaritsky} D.,  2013,
  \mn@doi [The Astrophysical Journal] {10.1088/0004-637X/778/1/14}, \href
  {https://ui.adsabs.harvard.edu/abs/2013ApJ...778...14G} {778, 14}

\bibitem[\protect\citeauthoryear{Haardt \& Madau}{Haardt \&
  Madau}{2001}]{haardt_modelling_2001}
Haardt F.,  Madau P.,  2001, in Neumann D.~M.,  Tran J. T.~V.,  eds, Clusters
  of {Galaxies} and the {High} {Redshift} {Universe} {Observed} in {X}-rays.

\bibitem[\protect\citeauthoryear{{Heitmann}, {Lawrence}, {Kwan}, {Habib}  \&
  {Higdon}}{{Heitmann} et~al.}{2014}]{heitmann_2014}
{Heitmann} K.,  {Lawrence} E.,  {Kwan} J.,  {Habib} S.,   {Higdon} D.,  2014,
  \mn@doi [The Astrophysical Journal] {10.1088/0004-637X/780/1/111}, \href
  {https://ui.adsabs.harvard.edu/abs/2014ApJ...780..111H} {780, 111}

\bibitem[\protect\citeauthoryear{{Heitmann} et~al.,}{{Heitmann}
  et~al.}{2016}]{heitmann_2016}
{Heitmann} K.,  et~al., 2016, \mn@doi [The Astrophysical Journal]
  {10.3847/0004-637X/820/2/108}, \href
  {https://ui.adsabs.harvard.edu/abs/2016ApJ...820..108H} {820, 108}

\bibitem[\protect\citeauthoryear{{Henden}, {Puchwein}, {Shen}  \&
  {Sijacki}}{{Henden} et~al.}{2018}]{henden_2018}
{Henden} N.~A.,  {Puchwein} E.,  {Shen} S.,   {Sijacki} D.,  2018, \mn@doi
  [Monthly Notices of the Royal Astronomical Society] {10.1093/mnras/sty1780},
  \href {https://ui.adsabs.harvard.edu/abs/2018MNRAS.479.5385H} {479, 5385}

\bibitem[\protect\citeauthoryear{{Henden}, {Puchwein}  \& {Sijacki}}{{Henden}
  et~al.}{2020}]{henden_2020}
{Henden} N.~A.,  {Puchwein} E.,   {Sijacki} D.,  2020, \mn@doi [Monthly Notices
  of the Royal Astronomical Society] {10.1093/mnras/staa2235}, \href
  {https://ui.adsabs.harvard.edu/abs/2020MNRAS.498.2114H} {498, 2114}

\bibitem[\protect\citeauthoryear{{Heymans} et~al.,}{{Heymans}
  et~al.}{2021}]{Heymans2021}
{Heymans} C.,  et~al., 2021, \mn@doi [\aap] {10.1051/0004-6361/202039063}, 646,
  A140

\bibitem[\protect\citeauthoryear{{Hinshaw} et~al.,}{{Hinshaw}
  et~al.}{2013}]{2013ApJS..208...19H}
{Hinshaw} G.,  et~al., 2013, \mn@doi [The Astrophysical Journal]
  {10.1088/0067-0049/208/2/19}, \href
  {https://ui.adsabs.harvard.edu/abs/2013ApJS..208...19H} {208, 19}

\bibitem[\protect\citeauthoryear{Hopkins}{Hopkins}{2013}]{hopkins_general_2013}
Hopkins P.~F.,  2013, \mn@doi [Monthly Notices of the Royal Astronomical
  Society] {10.1093/mnras/sts210}, 428, 2840

\bibitem[\protect\citeauthoryear{Kaiser}{Kaiser}{1987}]{kaiser_clustering_1987}
Kaiser N.,  1987, \mn@doi [Monthly Notices of the Royal Astronomical Society]
  {10.1093/mnras/227.1.1}, 227, 1

\bibitem[\protect\citeauthoryear{{Krause} et~al.,}{{Krause}
  et~al.}{2021}]{Krause2021arXiv}
{Krause} E.,  et~al., 2021, \mn@doi [arXiv e-prints]
  {10.48550/arXiv.2105.13548}, \href
  {https://ui.adsabs.harvard.edu/abs/2021arXiv210513548K} {p. arXiv:2105.13548}

\bibitem[\protect\citeauthoryear{{Lawrence} et~al.,}{{Lawrence}
  et~al.}{2017}]{lawrence_2017}
{Lawrence} E.,  et~al., 2017, \mn@doi [The Astrophysical Journal]
  {10.3847/1538-4357/aa86a9}, \href
  {https://ui.adsabs.harvard.edu/abs/2017ApJ...847...50L} {847, 50}

\bibitem[\protect\citeauthoryear{Le~Brun, McCarthy, Schaye  \& Ponman}{Le~Brun
  et~al.}{2014}]{le_brun_towards_2014}
Le~Brun A. M.~C.,  McCarthy I.~G.,  Schaye J.,   Ponman T.~J.,  2014, \mn@doi
  [Monthly Notices of the Royal Astronomical Society] {10.1093/mnras/stu608},
  441, 1270

\bibitem[\protect\citeauthoryear{Lewis, Challinor  \& Lasenby}{Lewis
  et~al.}{2000}]{lewis_efficient_2000}
Lewis A.,  Challinor A.,   Lasenby A.,  2000, \mn@doi [The Astrophysical
  Journal] {10.1086/309179}, 538, 473

\bibitem[\protect\citeauthoryear{{Lin}, {Stanford}, {Eisenhardt}, {Vikhlinin},
  {Maughan}  \& {Kravtsov}}{{Lin} et~al.}{2012}]{2012ApJ...745L...3L}
{Lin} Y.-T.,  {Stanford} S.~A.,  {Eisenhardt} P. R.~M.,  {Vikhlinin} A.,
  {Maughan} B.~J.,   {Kravtsov} A.,  2012, \mn@doi [The Astrophysical Journal]
  {10.1088/2041-8205/745/1/L3}, \href
  {https://ui.adsabs.harvard.edu/abs/2012ApJ...745L...3L} {745, L3}

\bibitem[\protect\citeauthoryear{{Lovisari}, {Reiprich}  \&
  {Schellenberger}}{{Lovisari} et~al.}{2015}]{2015A&A...573A.118L}
{Lovisari} L.,  {Reiprich} T.~H.,   {Schellenberger} G.,  2015, \mn@doi [\aap]
  {10.1051/0004-6361/201423954}, \href
  {https://ui.adsabs.harvard.edu/abs/2015A&A...573A.118L} {573, A118}

\bibitem[\protect\citeauthoryear{{Ma} \& {Fry}}{{Ma} \& {Fry}}{2000}]{ma_2000}
{Ma} C.-P.,  {Fry} J.~N.,  2000, \mn@doi [The Astrophysical Journal]
  {10.1086/312534}, \href
  {https://ui.adsabs.harvard.edu/abs/2000ApJ...531L..87M} {531, L87}

\bibitem[\protect\citeauthoryear{{Maughan}, {Jones}, {Forman}  \& {Van
  Speybroeck}}{{Maughan} et~al.}{2008}]{2008ApJS..174..117M}
{Maughan} B.~J.,  {Jones} C.,  {Forman} W.,   {Van Speybroeck} L.,  2008,
  \mn@doi [The Astrophysical Journal] {10.1086/521225}, \href
  {https://ui.adsabs.harvard.edu/abs/2008ApJS..174..117M} {174, 117}

\bibitem[\protect\citeauthoryear{McCarthy et~al.,}{McCarthy
  et~al.}{2010}]{mccarthy_case_2010}
McCarthy I.~G.,  et~al., 2010, \mn@doi [Monthly Notices of the Royal
  Astronomical Society] {10.1111/j.1365-2966.2010.16750.x}, 406, 822

\bibitem[\protect\citeauthoryear{McCarthy, Schaye, Bower, Ponman, Booth,
  Dalla~Vecchia  \& Springel}{McCarthy et~al.}{2011}]{mccarthy_gas_2011}
McCarthy I.~G.,  Schaye J.,  Bower R.~G.,  Ponman T.~J.,  Booth C.~M.,
  Dalla~Vecchia C.,   Springel V.,  2011, \mn@doi [Monthly Notices of the Royal
  Astronomical Society] {10.1111/j.1365-2966.2010.18033.x}, 412, 1965

\bibitem[\protect\citeauthoryear{McCarthy, Schaye, Bird  \& Le~Brun}{McCarthy
  et~al.}{2017}]{mccarthy_bahamas_2017}
McCarthy I.~G.,  Schaye J.,  Bird S.,   Le~Brun A. M.~C.,  2017, \mn@doi
  [Monthly Notices of the Royal Astronomical Society] {10.1093/mnras/stw2792},
  465, 2936

\bibitem[\protect\citeauthoryear{McCarthy, Bird, Schaye, Harnois-Deraps, Font
  \& van Waerbeke}{McCarthy et~al.}{2018}]{mccarthy_bahamas_2018}
McCarthy I.~G.,  Bird S.,  Schaye J.,  Harnois-Deraps J.,  Font A.~S.,   van
  Waerbeke L.,  2018, \mn@doi [Monthly Notices of the Royal Astronomical
  Society] {10.1093/mnras/sty377}, 476, 2999

\bibitem[\protect\citeauthoryear{{Mead}, {Heymans}, {Lombriser}, {Peacock},
  {Steele}  \& {Winther}}{{Mead} et~al.}{2016}]{Mead_16}
{Mead} A.~J.,  {Heymans} C.,  {Lombriser} L.,  {Peacock} J.~A.,  {Steele}
  O.~I.,   {Winther} H.~A.,  2016, \mn@doi [Monthly Notices of the Royal
  Astronomical Society] {10.1093/mnras/stw681}, \href
  {https://ui.adsabs.harvard.edu/abs/2016MNRAS.459.1468M} {459, 1468}

\bibitem[\protect\citeauthoryear{Mead, Tr{\"o}ster, Heymans, Van~Waerbeke  \&
  McCarthy}{Mead et~al.}{2020}]{mead_hydrodynamical_2020}
Mead A.~J.,  Tr{\"o}ster T.,  Heymans C.,  Van~Waerbeke L.,   McCarthy I.~G.,
  2020, \mn@doi [Astronomy and Astrophysics] {10.1051/0004-6361/202038308},
  641, A130

\bibitem[\protect\citeauthoryear{{Mead}, {Brieden}, {Tr{\"o}ster}  \&
  {Heymans}}{{Mead} et~al.}{2021}]{mead_hmcode-2020_2020}
{Mead} A.~J.,  {Brieden} S.,  {Tr{\"o}ster} T.,   {Heymans} C.,  2021, \mn@doi
  [Monthly Notices of the Royal Astronomical Society] {10.1093/mnras/stab082},
  \href {https://ui.adsabs.harvard.edu/abs/2021MNRAS.502.1401M} {502, 1401}

\bibitem[\protect\citeauthoryear{Moustakas et~al.,}{Moustakas
  et~al.}{2013}]{moustakas_primus:_2013}
Moustakas J.,  et~al., 2013, \mn@doi [The Astrophysical Journal]
  {10.1088/0004-637X/767/1/50}, 767, 50

\bibitem[\protect\citeauthoryear{{Mulroy} et~al.,}{{Mulroy}
  et~al.}{2019}]{Mulroy_2019}
{Mulroy} S.~L.,  et~al., 2019, \mn@doi [Monthly Notices of the Royal
  Astronomical Society] {10.1093/mnras/sty3484}, \href
  {https://ui.adsabs.harvard.edu/abs/2019MNRAS.484...60M} {484, 60}

\bibitem[\protect\citeauthoryear{Mummery, McCarthy, Bird  \& Schaye}{Mummery
  et~al.}{2017}]{mummery_separate_2017}
Mummery B.~O.,  McCarthy I.~G.,  Bird S.,   Schaye J.,  2017, \mn@doi [Monthly
  Notices of the Royal Astronomical Society] {10.1093/mnras/stx1469}, 471, 227

\bibitem[\protect\citeauthoryear{{Oppenheimer}, {Babul}, {Bah{\'e}}, {Butsky}
  \& {McCarthy}}{{Oppenheimer} et~al.}{2021}]{oppenheimer_2021}
{Oppenheimer} B.~D.,  {Babul} A.,  {Bah{\'e}} Y.,  {Butsky} I.~S.,   {McCarthy}
  I.~G.,  2021, \mn@doi [Universe] {10.3390/universe7070209}, \href
  {https://ui.adsabs.harvard.edu/abs/2021Univ....7..209O} {7, 209}

\bibitem[\protect\citeauthoryear{{Peacock} \& {Dodds}}{{Peacock} \&
  {Dodds}}{1994}]{Peacock_94}
{Peacock} J.~A.,  {Dodds} S.~J.,  1994, \mn@doi [Monthly Notices of the Royal
  Astronomical Society] {10.1093/mnras/267.4.1020}, \href
  {https://ui.adsabs.harvard.edu/abs/1994MNRAS.267.1020P} {267, 1020}

\bibitem[\protect\citeauthoryear{Peacock \& Smith}{Peacock \&
  Smith}{2000}]{peacock_halo_2000}
Peacock J.~A.,  Smith R.~E.,  2000, \mn@doi [Monthly Notices of the Royal
  Astronomical Society] {10.1046/j.1365-8711.2000.03779.x}, 318, 1144

\bibitem[\protect\citeauthoryear{{Pearson} et~al.,}{{Pearson}
  et~al.}{2017}]{2017MNRAS.469.3489P}
{Pearson} R.~J.,  et~al., 2017, \mn@doi [Monthly Notices of the Royal
  Astronomical Society] {10.1093/mnras/stx1081}, \href
  {https://ui.adsabs.harvard.edu/abs/2017MNRAS.469.3489P} {469, 3489}

\bibitem[\protect\citeauthoryear{Peebles}{Peebles}{1980}]{peebles_large-scale_1980}
Peebles P. J.~E.,  1980, The large-scale structure of the universe

\bibitem[\protect\citeauthoryear{{Pfeifer}, {McCarthy}, {Stafford}, {Brown},
  {Font}, {Kwan}, {Salcido}  \& {Schaye}}{{Pfeifer}
  et~al.}{2020}]{pfeifer_2020}
{Pfeifer} S.,  {McCarthy} I.~G.,  {Stafford} S.~G.,  {Brown} S.~T.,  {Font}
  A.~S.,  {Kwan} J.,  {Salcido} J.,   {Schaye} J.,  2020, \mn@doi [Monthly
  Notices of the Royal Astronomical Society] {10.1093/mnras/staa2240}, \href
  {https://ui.adsabs.harvard.edu/abs/2020MNRAS.498.1576P} {498, 1576}

\bibitem[\protect\citeauthoryear{{Planck Collaboration} et~al.,}{{Planck
  Collaboration} et~al.}{2020}]{planck_2018_cosmology}
{Planck Collaboration} et~al., 2020, \mn@doi [\aap]
  {10.1051/0004-6361/201833910}, \href
  {https://ui.adsabs.harvard.edu/abs/2020A&A...641A...6P} {641, A6}

\bibitem[\protect\citeauthoryear{{Planelles}, {Borgani}, {Fabjan}, {Killedar},
  {Murante}, {Granato}, {Ragone-Figueroa}  \& {Dolag}}{{Planelles}
  et~al.}{2014}]{planelles_2014}
{Planelles} S.,  {Borgani} S.,  {Fabjan} D.,  {Killedar} M.,  {Murante} G.,
  {Granato} G.~L.,  {Ragone-Figueroa} C.,   {Dolag} K.,  2014, \mn@doi [Monthly
  Notices of the Royal Astronomical Society] {10.1093/mnras/stt2141}, \href
  {https://ui.adsabs.harvard.edu/abs/2014MNRAS.438..195P} {438, 195}

\bibitem[\protect\citeauthoryear{{Pratt}, {Croston}, {Arnaud}  \&
  {B{\"o}hringer}}{{Pratt} et~al.}{2009}]{2009A&A...498..361P}
{Pratt} G.~W.,  {Croston} J.~H.,  {Arnaud} M.,   {B{\"o}hringer} H.,  2009,
  \mn@doi [\aap] {10.1051/0004-6361/200810994}, \href
  {https://ui.adsabs.harvard.edu/abs/2009A&A...498..361P} {498, 361}

\bibitem[\protect\citeauthoryear{Price}{Price}{2008}]{price_modelling_2008}
Price D.~J.,  2008, \mn@doi [Journal of Computational Physics]
  {10.1016/j.jcp.2008.08.011}, 227, 10040

\bibitem[\protect\citeauthoryear{{Puchwein}, {Sijacki}  \&
  {Springel}}{{Puchwein} et~al.}{2008}]{puchwein_2008}
{Puchwein} E.,  {Sijacki} D.,   {Springel} V.,  2008, \mn@doi [The
  Astrophysical Journal] {10.1086/593352}, \href
  {https://ui.adsabs.harvard.edu/abs/2008ApJ...687L..53P} {687, L53}

\bibitem[\protect\citeauthoryear{{Rasmussen} \& {Ponman}}{{Rasmussen} \&
  {Ponman}}{2009}]{2009MNRAS.399..239R}
{Rasmussen} J.,  {Ponman} T.~J.,  2009, \mn@doi [Monthly Notices of the Royal
  Astronomical Society] {10.1111/j.1365-2966.2009.15244.x}, \href
  {https://ui.adsabs.harvard.edu/abs/2009MNRAS.399..239R} {399, 239}

\bibitem[\protect\citeauthoryear{{Sanderson}, {O'Sullivan}, {Ponman},
  {Gonzalez}, {Sivanandam}, {Zabludoff}  \& {Zaritsky}}{{Sanderson}
  et~al.}{2013}]{2013MNRAS.429.3288S}
{Sanderson} A. J.~R.,  {O'Sullivan} E.,  {Ponman} T.~J.,  {Gonzalez} A.~H.,
  {Sivanandam} S.,  {Zabludoff} A.~I.,   {Zaritsky} D.,  2013, \mn@doi [Monthly
  Notices of the Royal Astronomical Society] {10.1093/mnras/sts586}, \href
  {https://ui.adsabs.harvard.edu/abs/2013MNRAS.429.3288S} {429, 3288}

\bibitem[\protect\citeauthoryear{Schaye}{Schaye}{2004}]{schaye_star_2004}
Schaye J.,  2004, \mn@doi [The Astrophysical Journal] {10.1086/421232}, 609,
  667

\bibitem[\protect\citeauthoryear{Schaye \& Dalla~Vecchia}{Schaye \&
  Dalla~Vecchia}{2008}]{schaye_relation_2008}
Schaye J.,  Dalla~Vecchia C.,  2008, \mn@doi [Monthly Notices of the Royal
  Astronomical Society] {10.1111/j.1365-2966.2007.12639.x}, 383, 1210

\bibitem[\protect\citeauthoryear{Schaye et~al.,}{Schaye
  et~al.}{2010}]{schaye_physics_2010}
Schaye J.,  et~al., 2010, \mn@doi [Monthly Notices of the Royal Astronomical
  Society] {10.1111/j.1365-2966.2009.16029.x}, 402, 1536

\bibitem[\protect\citeauthoryear{Schaye et~al.,}{Schaye
  et~al.}{2015}]{schaye_eagle_2015}
Schaye J.,  et~al., 2015, \mn@doi [Monthly Notices of the Royal Astronomical
  Society] {10.1093/mnras/stu2058}, 446, 521

\bibitem[\protect\citeauthoryear{Schneider \& Teyssier}{Schneider \&
  Teyssier}{2015}]{schneider_new_2015}
Schneider A.,  Teyssier R.,  2015, \mn@doi [Journal of Cosmology and
  Astroparticle Physics] {10.1088/1475-7516/2015/12/049}, 2015, 049

\bibitem[\protect\citeauthoryear{{Schneider}, {Stoira}, {Refregier}, {Weiss},
  {Knabenhans}, {Stadel}  \& {Teyssier}}{{Schneider}
  et~al.}{2020}]{schneider_2020}
{Schneider} A.,  {Stoira} N.,  {Refregier} A.,  {Weiss} A.~J.,  {Knabenhans}
  M.,  {Stadel} J.,   {Teyssier} R.,  2020, \mn@doi [\jcap]
  {10.1088/1475-7516/2020/04/019}, \href
  {https://ui.adsabs.harvard.edu/abs/2020JCAP...04..019S} {2020, 019}

\bibitem[\protect\citeauthoryear{{Secco} et~al.,}{{Secco}
  et~al.}{2022}]{Secco2022}
{Secco} L.~F.,  et~al., 2022, \mn@doi [\prd] {10.1103/PhysRevD.105.023515},
  \href {https://ui.adsabs.harvard.edu/abs/2022PhRvD.105b3515S} {105, 023515}

\bibitem[\protect\citeauthoryear{Seljak}{Seljak}{2000}]{seljak_analytic_2000}
Seljak U.,  2000, \mn@doi [Monthly Notices of the Royal Astronomical Society]
  {10.1046/j.1365-8711.2000.03715.x}, 318, 203

\bibitem[\protect\citeauthoryear{Semboloni, Hoekstra, Schaye, van Daalen  \&
  McCarthy}{Semboloni et~al.}{2011}]{semboloni_quantifying_2011}
Semboloni E.,  Hoekstra H.,  Schaye J.,  van Daalen M.~P.,   McCarthy I.~G.,
  2011, \mn@doi [Monthly Notices of the Royal Astronomical Society]
  {10.1111/j.1365-2966.2011.19385.x}, 417, 2020

\bibitem[\protect\citeauthoryear{{Sereno} et~al.,}{{Sereno}
  et~al.}{2020}]{2020MNRAS.492.4528S}
{Sereno} M.,  et~al., 2020, \mn@doi [Monthly Notices of the Royal Astronomical
  Society] {10.1093/mnras/stz3425}, \href
  {https://ui.adsabs.harvard.edu/abs/2020MNRAS.492.4528S} {492, 4528}

\bibitem[\protect\citeauthoryear{Smith et~al.,}{Smith
  et~al.}{2003}]{smith_stable_2003}
Smith R.~E.,  et~al., 2003, \mn@doi [Monthly Notices of the Royal Astronomical
  Society] {10.1046/j.1365-8711.2003.06503.x}, 341, 1311

\bibitem[\protect\citeauthoryear{Springel}{Springel}{2005}]{springel_cosmological_2005}
Springel V.,  2005, \mn@doi [Monthly Notices of the Royal Astronomical Society]
  {10.1111/j.1365-2966.2005.09655.x}, 364, 1105

\bibitem[\protect\citeauthoryear{Springel, Di~Matteo  \& Hernquist}{Springel
  et~al.}{2005}]{springel_modelling_2005}
Springel V.,  Di~Matteo T.,   Hernquist L.,  2005, \mn@doi [Monthly Notices of
  the Royal Astronomical Society] {10.1111/j.1365-2966.2005.09238.x}, 361, 776

\bibitem[\protect\citeauthoryear{Springel et~al.,}{Springel
  et~al.}{2018}]{springel_first_2018}
Springel V.,  et~al., 2018, \mn@doi [Monthly Notices of the Royal Astronomical
  Society] {10.1093/mnras/stx3304}, 475, 676

\bibitem[\protect\citeauthoryear{{Stafford}, {McCarthy}, {Crain}, {Salcido},
  {Schaye}, {Font}, {Kwan}  \& {Pfeifer}}{{Stafford}
  et~al.}{2020}]{stafford_2020}
{Stafford} S.~G.,  {McCarthy} I.~G.,  {Crain} R.~A.,  {Salcido} J.,  {Schaye}
  J.,  {Font} A.~S.,  {Kwan} J.,   {Pfeifer} S.,  2020, \mn@doi [Monthly
  Notices of the Royal Astronomical Society] {10.1093/mnras/staa129}, \href
  {https://ui.adsabs.harvard.edu/abs/2020MNRAS.493..676S} {493, 676}

\bibitem[\protect\citeauthoryear{{Sun}, {Voit}, {Donahue}, {Jones}, {Forman}
  \& {Vikhlinin}}{{Sun} et~al.}{2009}]{2009ApJ...693.1142S}
{Sun} M.,  {Voit} G.~M.,  {Donahue} M.,  {Jones} C.,  {Forman} W.,
  {Vikhlinin} A.,  2009, \mn@doi [The Astrophysical Journal]
  {10.1088/0004-637X/693/2/1142}, \href
  {https://ui.adsabs.harvard.edu/abs/2009ApJ...693.1142S} {693, 1142}

\bibitem[\protect\citeauthoryear{{Takahashi}, {Sato}, {Nishimichi}, {Taruya}
  \& {Oguri}}{{Takahashi} et~al.}{2012}]{Takahashi_12}
{Takahashi} R.,  {Sato} M.,  {Nishimichi} T.,  {Taruya} A.,   {Oguri} M.,
  2012, \mn@doi [The Astrophysical Journal] {10.1088/0004-637X/761/2/152},
  \href {https://ui.adsabs.harvard.edu/abs/2012ApJ...761..152T} {761, 152}

\bibitem[\protect\citeauthoryear{{Tr{\"o}ster} et~al.,}{{Tr{\"o}ster}
  et~al.}{2021}]{Troster2021}
{Tr{\"o}ster} T.,  et~al., 2021, \mn@doi [\aap] {10.1051/0004-6361/202039805},
  649, A88

\bibitem[\protect\citeauthoryear{{Tr{\"o}ster} et~al.,}{{Tr{\"o}ster}
  et~al.}{2022}]{troster_2022}
{Tr{\"o}ster} T.,  et~al., 2022, \mn@doi [\aap] {10.1051/0004-6361/202142197},
  \href {https://ui.adsabs.harvard.edu/abs/2022A&A...660A..27T} {660, A27}

\bibitem[\protect\citeauthoryear{Valkenburg \& Villaescusa-Navarro}{Valkenburg
  \& Villaescusa-Navarro}{2017}]{valkenburg_accurate_2017}
Valkenburg W.,  Villaescusa-Navarro F.,  2017, \mn@doi [Monthly Notices of the
  Royal Astronomical Society] {10.1093/mnras/stx376}, 467, 4401

\bibitem[\protect\citeauthoryear{{Vikhlinin}, {Kravtsov}, {Forman}, {Jones},
  {Markevitch}, {Murray}  \& {Van Speybroeck}}{{Vikhlinin}
  et~al.}{2006}]{2006ApJ...640..691V}
{Vikhlinin} A.,  {Kravtsov} A.,  {Forman} W.,  {Jones} C.,  {Markevitch} M.,
  {Murray} S.~S.,   {Van Speybroeck} L.,  2006, \mn@doi [The Astrophysical
  Journal] {10.1086/500288}, \href
  {https://ui.adsabs.harvard.edu/abs/2006ApJ...640..691V} {640, 691}

\bibitem[\protect\citeauthoryear{Wendland}{Wendland}{1995}]{wendland_piecewise_1995}
Wendland H.,  1995, \mn@doi [Advances in Computational Mathematics]
  {10.1007/BF02123482}, 4, 389

\bibitem[\protect\citeauthoryear{{White}, {Navarro}, {Evrard}  \&
  {Frenk}}{{White} et~al.}{1993}]{white_1993}
{White} S. D.~M.,  {Navarro} J.~F.,  {Evrard} A.~E.,   {Frenk} C.~S.,  1993,
  \mn@doi [\nat] {10.1038/366429a0}, \href
  {https://ui.adsabs.harvard.edu/abs/1993Natur.366..429W} {366, 429}

\bibitem[\protect\citeauthoryear{Wiersma, Schaye  \& Smith}{Wiersma
  et~al.}{2009a}]{wiersma_effect_2009}
Wiersma R. P.~C.,  Schaye J.,   Smith B.~D.,  2009a, \mn@doi [Monthly Notices
  of the Royal Astronomical Society] {10.1111/j.1365-2966.2008.14191.x}, 393,
  99

\bibitem[\protect\citeauthoryear{Wiersma, Schaye, Theuns, Dalla~Vecchia  \&
  Tornatore}{Wiersma et~al.}{2009b}]{wiersma_chemical_2009}
Wiersma R. P.~C.,  Schaye J.,  Theuns T.,  Dalla~Vecchia C.,   Tornatore L.,
  2009b, \mn@doi [Monthly Notices of the Royal Astronomical Society]
  {10.1111/j.1365-2966.2009.15331.x}, 399, 574

\bibitem[\protect\citeauthoryear{van Daalen \& Schaye}{van Daalen \&
  Schaye}{2015}]{van_daalen_contributions_2015}
van Daalen M.~P.,  Schaye J.,  2015, \mn@doi [Monthly Notices of the Royal
  Astronomical Society] {10.1093/mnras/stv1456}, 452, 2247

\bibitem[\protect\citeauthoryear{van Daalen, Schaye, Booth  \&
  Dalla~Vecchia}{van Daalen et~al.}{2011}]{van_daalen_effects_2011}
van Daalen M.~P.,  Schaye J.,  Booth C.~M.,   Dalla~Vecchia C.,  2011, \mn@doi
  [Monthly Notices of the Royal Astronomical Society]
  {10.1111/j.1365-2966.2011.18981.x}, 415, 3649

\bibitem[\protect\citeauthoryear{{van Daalen}, {McCarthy}  \& {Schaye}}{{van
  Daalen} et~al.}{2020}]{van_daalen_2020}
{van Daalen} M.~P.,  {McCarthy} I.~G.,   {Schaye} J.,  2020, \mn@doi [Monthly
  Notices of the Royal Astronomical Society] {10.1093/mnras/stz3199}, \href
  {https://ui.adsabs.harvard.edu/abs/2020MNRAS.491.2424V} {491, 2424}

\makeatother
\end{thebibliography}

%\begin{thebibliography}{99}
%\end{thebibliography}

%%%%%%%%%%%%%%%%%%%%%%%%%%%%%%%%%%%%%%%%%%%%%%%%%%

%%%%%%%%%%%%%%%%% APPENDICES %%%%%%%%%%%%%%%%%%%%%

%%%%%%%%%%%%%%%%%%%%%%%%%%%%%%%%%%%%%%%%%%%%%%%%%%

% Don't change these lines
\bsp	% typesetting comment
\label{lastpage}
\end{document}